%% file: paper.tex
\documentclass[conference,compsoc]{./IEEEtran}
\usepackage[normalem]{ulem}
\usepackage{xspace}
\usepackage[numbers,sort&compress]{natbib}
\usepackage{tikz}
\usepackage[hyphens]{url}
\usepackage[hyphenbreaks]{breakurl}
\usepackage{hyperref}
\usepackage{fancyvrb}

\usepackage{tgcursor}
\usepackage{pifont}
\usepackage{multicol}
\usepackage{cleveref}
\usepackage{subfig}
\usepackage{multirow}
\usepackage{paralist}
\usepackage{adjustbox}
\usepackage{array}
\usepackage{wasysym}
\usepackage{amssymb}
\usepackage[warn]{textcomp}

\definecolor{LinkColor}{rgb}{0.0, 0.08, 0.66}
\hypersetup{
    colorlinks=true,
    linkcolor=LinkColor, % color for internal links
    urlcolor=LinkColor, % color of URLs
    citecolor=black, % color of cites
    anchorcolor=black % target text, we don't want a link there
}

\usepackage[subtle,floats,paragraphs=normal,indent=normal,title=normal,bibbreaks=normal,
            mathspacing=normal,wordspacing=normal,tracking=normal]{savetrees}

\usepackage[hang,flushmargin]{footmisc}

\usepackage{soul}
\setul{2pt}{.4pt}% 1pt below contents

\usepackage{enumitem}

\usepackage{color}
\usepackage{tcolorbox}

\usepackage{xcolor}

\newcommand{\overhead}[1]{%
\begin{tikzpicture}
 \draw (0,0) circle (0.8ex);\fill (0.8ex,0) arc (0:#1:0.8ex) -- (0,0) -- cycle;
\end{tikzpicture}%
}

\newcommand\customparheader[1]{{\emph{\textbf{#1}}}}
\newcommand\customparheaderb[1]{{\textbf{#1}}}

\newcommand\smlcite[1]{{\tiny \cite{#1}}}

\newcommand\customtt[1]{{\small \texttt{#1}}}
\newcommand\tinytt[1]{{\scriptsize \texttt{#1}}}

\newcommand{\nooverhead}[1][]{\overhead{0}}
\newcommand{\lowoverhead}[1][]{\overhead{90}}
\newcommand{\mediumoverhead}[1][]{\overhead{180}}
\newcommand{\highoverhead}[1][]{\overhead{270}}
\newcommand{\veryhighoverhead}[1][]{\overhead{360}}

\newcommand{\manual}[1][]{\overhead{0}}
\newcommand{\lowautomation}[1][]{\overhead{90}}
\newcommand{\mediumautomation}[1][]{\overhead{180}}
\newcommand{\highautomation}[1][]{\overhead{270}}
\newcommand{\fullauto}[1][]{\overhead{360}}

\newcommand*{\eg}{e.g.,\@\xspace}
\newcommand*{\ie}{i.e.,\@\xspace}
\newcommand*{\vs}{vs.\@\xspace}
\newcommand*{\cf}{cf.\@\xspace}

\newcommand*{\pone}{{\small P1}\@\xspace}
\newcommand*{\ptwo}{{\small P2}\@\xspace}
\newcommand*{\pthree}{{\small P3}\@\xspace}
\newcommand*{\ponetotwo}{{\small P1-P2}\@\xspace}
\newcommand*{\ponetothree}{{\small P1-P3}\@\xspace}
\newcommand*{\ponetothreeverbose}{{\small P1, P2,} and {\small P3}\@\xspace}

\newcommand*{\totalpapers}{9,083\xspace}
\newcommand*{\potentiallyrelated}{923\xspace}
\newcommand*{\totalinstudy}{211\xspace}
\newcommand*{\systematicsearch}{96\xspace}
\newcommand*{\recsearch}{66\xspace}
\newcommand*{\ownexpertise}{49\xspace}
\newcommand*{\intables}{106\xspace}

\newcommand*{\popconpackages}{1,520\xspace}
\newcommand*{\totalfound}{61\xspace}
\newcommand*{\classesnum}{13\xspace}
\newcommand*{\totalsystemfound}{56\xspace}
\newcommand*{\systemfound}{19\xspace}
\newcommand*{\systemfoundvendor}{37\xspace}
\newcommand*{\rworksfound}{3\xspace}
\newcommand*{\expertfound}{2\xspace}

\definecolor{beaublue}{HTML}{e2e6ed}
\definecolor{lightgrey}{RGB}{244,244,244}
\definecolor{darkgrey}{RGB}{99,99,99}

\tcbset{
	boxrule=0.5pt,
	boxsep=0.5pt,
	left=3pt,
	right=3pt,
	top=3pt,
	bottom=3pt,
	sharp corners,
	colframe=white,
}

\newcommand{\greyblock}[3]{
	\begin{center}
	\begin{tcolorbox}[width=\linewidth, colback=#1]
		\textbf{#2:} #3
	\end{tcolorbox}
	\end{center}
}

\newcounter{insights}
\newcommand{\insight}[1]{
	\greyblock{lightgrey}{Insight~\number\value{insights}}{#1}%
	\stepcounter{insights}
}

\newcounter{definitions}
\newcommand{\definition}[1]{
	\greyblock{beaublue}{Definition~\MakeUppercase{\romannumeral\value{definitions}}}{#1}%
	\stepcounter{definitions}
}

\begin{document}

\newcommand{\asterisk}{1}
\newcommand{\ubc}{$\dagger$}
\newcommand{\sci}{{\tiny$\infty$}}
\newcommand{\serenitix}{{$\ddagger$}}
\newcommand{\manchester}{\tiny$\nabla$}

\author{\IEEEauthorblockN{Hugo Lefeuvre\textsuperscript{\ubc\asterisk}, Nathan Dautenhahn\textsuperscript{\serenitix\asterisk}, David Chisnall\textsuperscript{\sci}, Pierre Olivier\textsuperscript{\manchester}}
\IEEEauthorblockA{\textit{\textsuperscript{\ubc}The University of British Columbia, \textsuperscript{\serenitix}Serenitix, \textsuperscript{\sci}SCI Semiconductor, \textsuperscript{\manchester}The University of Manchester}}}

% Reduce the spacing before \paragraph a little bit
\makeatletter
\renewcommand\paragraph{\@startsection{paragraph}{4}{\z@}%
                                    {1.8ex \@plus1ex \@minus.2ex}%
                                    {-1em}%
                                    {\normalfont\normalsize\bfseries}}
\makeatother

\title{SoK: Software Compartmentalization}

\date{}
\maketitle

\thispagestyle{empty}

\newcommand{\customfootnotetext}[2]{{% Group to localize change to footnote
  \renewcommand{\thefootnote}{#1}% Update footnote counter representation
  \footnotetext[0]{#2}}}% Print footnote text

\customfootnotetext{\asterisk}{This work was primarily done while Hugo Lefeuvre
was affiliated with the University of Manchester and SCI Semiconductor, and
Nathan Dautenhahn with Rice University (with minor contributions at Riverside
Research).}

\input{00-abstract}
\input{01-introduction}
\input{02-background}
\input{03-approaches}
\input{04-deployed}
\input{05-limitations}
\input{06-relatedworks}
\input{07-conclusion}

\section*{Acknowledgments}

We are grateful to the anonymous reviewers and to our anonymous shepherd, as
well as to Peter Pietzuch, Mikel Luján, Shravan Narayan, Sid Agrawal, and
Aastha Mehta for their insightful feedback. This work was partly funded by a
studentship from NEC Labs Europe, a Microsoft Research PhD Fellowship, the
EPSRC grants EP/V012134/1 (UniFaaS), EP/V000225/1 (SCorCH), and EP/X015610/1
(FlexCap), and the NSF CNS grants \#2008867, \#2146537.

% trick to reduce the size of the bibliography (capped at 5 pages)
\let\oldbibliography\thebibliography
\renewcommand{\thebibliography}[1]{%
  \oldbibliography{#1}%
  \setlength{\itemsep}{0.4pt}%
}

\bibliographystyle{./custom}
{
\newcommand\biburl[1]{{\scriptsize \url{#1}}}
\newcommand\compressabit[1]{{\fontdimen4\font=0.2em #1}}
\newcommand\compressmore[1]{{\fontdimen4\font=0.4em #1}}
\footnotesize
\bibliography{paper}
}

%\appendix
%\label{appendix}

%\input{A-definitions}
\input{B-methodology}
\input{C-methodology}

\section*{A3. Meta-Review}

The following meta-review was prepared by the program committee for the 2025
IEEE Symposium on Security and Privacy (S\&P) as part of the review process as
detailed in the call for papers.

\subsection{Summary}

This paper investigates why the compartmentalization of software is still not a
mainstream practice and how this status quo can be improved. The paper proposes
a unified model for the systematic analysis, comparison, and directing of
compartmentalization approaches. Additionally, the paper analyzes how
compartmentalization is adopted in real-world applications and presents some
recommendations for the research and adoption of compartmentalization.

\subsection{Scientific Contributions}
\begin{itemize}
\item Independent Confirmation of Important Results with Limited Prior Research
\item Provides a Valuable Step Forward in an Established Field
\end{itemize}

\subsection{Reasons for Acceptance}
\begin{enumerate}
\item A comprehensive taxonomy of existing efforts with a structured approach to categorizing compartmentalization methods across multiple dimensions
\item A large-scale review of both research efforts and mainstream software systems
\end{enumerate}

\end{document}

%% file: 00-abstract.tex
\begin{abstract}
Decomposing large systems into smaller components with limited privileges has long been recognized as an effective means to minimize the impact of exploits.
Despite historical roots, demonstrated benefits, and a plethora of research efforts in academia and industry, the compartmentalization of software is still not a mainstream practice.
This paper investigates why, and how this status quo can be improved.
Noting that existing approaches are fraught with inconsistencies in terminology and analytical methods, we propose a unified model for the systematic analysis, comparison, and directing of compartmentalization approaches.
We use this model to review \totalinstudy{} research efforts and analyze \totalfound{} mainstream compartmentalized systems, confronting them to understand the limitations of both research and production works.
Among others, our findings reveal that mainstream efforts largely rely on manual methods, custom abstractions, and legacy mechanisms, poles apart from recent research.
We conclude with recommendations: compartmentalization should be solved holistically; progress is needed towards simplifying the definition of compartmentalization policies; towards better challenging our threat models in the light of confused deputies and hardware limitations; as well as towards bridging the gaps we pinpoint between research and mainstream needs.
This paper not only maps the historical and current landscape of compartmentalization, but also sets forth a framework to foster their evolution and adoption.
\end{abstract}

%% file: 01-introduction.tex
\section{Introduction}
\label{sec:introduction}

Despite decades of effort, vulnerabilities still plague software, and thwarting them remains a game of cat and mouse.
The \emph{principle of least privilege} (PoLP)~\cite{Saltzer1975} is the last line of defense when protections fail or when flaws are unknown.
By granting each entity only the privileges needed, the PoLP ensures that a compromise of one part will not imply that of the whole.
\emph{Software compartmentalization} is a prominent implementation of the
PoLP, in which developers divide a
large program into smaller, lesser privileged components, to reduce the impact of potential security breaches.

Software compartmentalization inherits from a large body of work, starting with processes~\cite{Daley1968}; including OS models such as microkernels~\cite{Young1987, Liedtke1995, Herder2006, Klein2009, Elphinstone2013}, security and separation kernels~\cite{Rushby1981, Rushby1982, Ames1983, Whitaker2003, AlvesFoss2006}, or capability OSes~\cite{Heiser1999, Chase1994}; all the way to fine-grain application compartmentalization in the 2000s~\cite{Wahbe1993, Kilpatrick2003, VahldiekOberwagner2019, Narayan2020} following qmail~\cite{Bernstein2007}, Postfix~\cite{Hafiz2008}, or OpenSSH~\cite{Provos2003}.
Its promises are plenty: containing memory safety issues~\cite{BoydWickizer2010, Roessler2021}, untrusted third parties~\cite{Acar2016, Narayan2020}, or unsafe parts of safe languages~\cite{Mergendahl2022, Sung2020, Rivera2021, Kirth2022, Bang2023}, isolating cryptographic secrets~\cite{Litton2016, Liu2015, VahldiekOberwagner2019} or shadow stacks~\cite{Burow2019}, thwarting supply-chain attacks~\cite{Vasilakis2018, Ghosn2021} and side-channels~\cite{Jenkins2020, Narayan2021, Milburn2022}, or providing fault resilience~\cite{Liedtke1995, Narayanan2020a}.

Despite longstanding recognition within the academic sphere and proven effectiveness in seminal industry projects, the adoption of compartmentalization techniques in mainstream software remains inconsistent: compartmentalizing is still far from being a common engineering practice.
Take as example the isolation of cryptographic secrets, a practice long advocated by the community~\cite{Jin2022, Gu2022, Sartakov2021, Gudka2015, Bauer2021, Litton2016, VahldiekOberwagner2019, Brumley2004, Chen2016, Bittau2008, Strackx2015, Guan2015, Lee2018, Lind2017} but without viable adoption by leading cryptography libraries.
At a time of growing threats, we are missing out on the security benefits compartmentalization can bring.
This paper investigates the reasons behind this status quo, and how to improve it.

Research speculated that this situation is due to a lack of automation~\cite{Wu2013, Bauer2021}, limitations of mechanisms~\cite{Park2019, Yuan2023}, excessive performance overheads~\cite{VahldiekOberwagner2019, Norton2016}, a lack of strong security guarantees~\cite{Lefeuvre2023, Hu2015}, among others.
All are likely part of the problem, and through decades of research progress was made on every front.
Today the community lacks a global overview of this progress.
Designing for or retrofitting compartmentalization is often hampered by inconsistencies in the understanding and application of its concepts.
Existing models and terminologies are numerous, ad-hoc, and often contradictory, leading to confusion and a growing body of work that cannot be compared.
The lack of a systematic perspective leads to a mismatch between what software compartmentalization needs to progress towards the mainstream, and the focus and framing of research efforts: most do not tackle compartmentalization's key aspects as a whole and thus produce solutions that cannot be relevant to making it a mainstream practice.

Recognizing these challenges, we propose a unified model providing a consistent framework for defining, understanding, and implementing compartmentalization.
This model comes with a comprehensive taxonomy that allows to classify compartmentalization strategies based on their policy definition methods, abstractions, and mechanisms, providing a basis for systematic evaluation and comparison.

We validate our model and taxonomy by systematizing \totalinstudy{} research and \totalfound{} mainstream software systems implementing compartmentalization.
Doing so, this study provides unique insights into where mainstream efforts have fallen behind, and which problems research needs to focus on to reach the mainstream.
We show how modern production software, if at all compartmentalizing, still vastly relies on manual separation, custom abstractions, and heavyweight legacy
mechanisms, failing to adapt to the evolving security landscape.
For future research, we stress the need for a more holistic approach to compartmentalization, the simplification of policy definition, stronger and more holistic threat models in the light of confused deputies and hardware limitations, as well as more attention to the gaps we highlight between research and production. Overall, this SoK contributes:
\begin{itemize}[leftmargin=0.5cm,noitemsep]
\item A fundamental model and conceptual framework for compartmentalization (\S\ref{sec:model}).

\item A taxonomy for evaluating compartmentalization approaches and its systematic application to a wide set of research (\S\ref{sec:stateoftheart}) and mainstream (\S\ref{sec:deployed}) efforts.

\item Insights throughout \S\ref{sec:background}, \S\ref{sec:stateoftheart} and \S\ref{sec:deployed} and their consolidation into high-level challenges (\S\ref{sec:limitations}) that compartmentalization research should focus on to foster mainstream adoption.

\end{itemize}

%% file: 02-background.tex
\vspace{-0.1cm}
\section{A Model of Software Compartmentalization}\label{sec:background}\label{sec:model}
\vspace{-0.1cm}

\label{def-subject}
\label{def-pdomain}
Early seminal works introduced models such as hierarchical layers and
separation of concerns~\cite{Dennis1966, Dijkstra1968, Hansen1970,
Anderson1972}, the object capability model~\cite{Dennis1966, Lampson1974}, the
access control matrix~\cite{Lampson1974}, the principle of
least privilege~\cite{Saltzer1974, Saltzer1975}, information
hiding~\cite{Parnas1972}, or information-flow control~\cite{Schell1973,
Walter1975, Denning1976}.  Although these models deeply influenced modern
compartmentalization, we observe that use-cases, practices, and enforcement
means have evolved such that these seminal works have become non-trivial to map
to modern compartmentalization approaches.  Thus, to consistently characterize
compartmentalization, we must first define and model it.

In the following, we adopt definitions of \emph{subject}, \emph{object}, and
\emph{permission} from Saltzer and Schroeder~\cite{Saltzer1975} and
Miller~\cite{Miller2006}, which we summarize as follows: a \ul{subject} (or
\emph{principal}~\cite{Saltzer1975}) is a unit of computation (\eg an assembly
instruction, a thread of execution), an \ul{object} is a unit of privilege
enforcement (\eg a byte of memory, a socket), and \ul{permissions} are actions
subjects may perform on objects (\eg read, write).  We refer to a maximal set
of subjects sharing identical (sets of) permissions as a \emph{protection
domain}.\footnote{Note that defining \emph{protection domain} from the
viewpoint of subjects and their permissions and not from that of objects like
Saltzer~\cite{Saltzer1975} is deliberate to better match modern practices where
domains are code-centric (see \S\ref{sec:stateoftheart}). Nevertheless, both
definitions can be used interchangeably.} Finally, we define a \emph{program}
similarly to ISO/IEC~\cite{ISO_PROGRAM_DEF} as a syntactic unit (conforming to
the rules of one or more programming languages), composed of subjects and
objects, needed to solve a certain function, task, or problem.  This allows us
to define software compartmentalization as:

\label{def-comp}
\definition{
\hypertarget{swcomp}{A \emph{compartmentalization}} of a program $P$
is the set of (1) a policy to separate $P$ into two or more protection domains
(called \emph{compartments}), and (2) the enforcement of this policy at runtime.
}

Compartmentalization can be applied to any program: applications, OS kernels
(microkernels~\cite{Liedtke1995} are compartmentalized kernels),
hypervisors~\cite{Shi2017}, firmware~\cite{Khan2023}, among others.
Compartmentalization can be \emph{retrofitted} into existing monolithic
programs, or present in the initial design of a program. This definition is
similar to Provos et al.~\cite{Provos2003}'s \emph{privilege separation},
expanded to general program types and trust models,\footnote{The definition of
privilege separation by Provos et al.~\cite{Provos2003} separates
\emph{applications} with a \emph{sandbox} threat model, which constitutes a
subset of compartmentalization as we define it in Definition
\hyperlink{swcomp}{I}.} and corresponds to the application of the principle of
least privilege~\cite{Saltzer1975} within a single program.

A key problem of compartmentalization is validating all control and data flows
at compartment boundaries~\cite{Provos2003}; we refer to this as ensuring
\hypertarget{isafety}{\emph{interface safety}}.  Improper validation results,
among others, in confused deputy vulnerabilities~\cite{Hardy1988} well-studied
in previous works~\cite{Lefeuvre2023, Chien2023, Narayan2020, Hu2015,
Checkoway2013}.  Interface safety is the instantiation in compartmentalization
of the problem of information-flow control~\cite{Denning1976}.

\vspace{-0.1cm}
\paragraph{Evaluating Compartmentalization}
\label{subsec:compartmentalization-goals}
Compartmentalization approaches strive to optimize some or all of four
properties: the \emph{security and safety} benefits of
compartmentalization (properties enforced, interface safety guarantees); the
\emph{performance} of separated software (compared to a monolithic
design); the \emph{compatibility} of compartmentalization with existing
software and programming idioms (\eg to minimize the reimplementation effort
needed by programmers to compartmentalize); and the \emph{usability} of
separated software, ensuring that non-expert end-users can correctly operate
(\eg configure, maintain, monitor) compartmentalized software.

Striking the right balance between these properties is key to match
compartmentalization approaches with real-world uses.  There is no silver
bullet~\cite{Parmer2007, Parmer2012, Gudka2015, Roessler2021, Dong2013,
Lefeuvre2022}; all approaches discussed next adopt different points in the
design space.  These trade-offs are thus central to this SoK.

\vspace{-0.1cm}
\paragraph{Scope of this SoK}

As per Definition \hyperlink{swcomp}{I}, we refer to software
compartmentalization as applied \emph{within} one program (application, kernel,
etc.), similarly to Provos~\cite{Provos2003}'s privilege separation.  As we
show throughout this paper, this covers a vast and coherent body of work.
Other isolation techniques~\cite{Shu2016} \emph{across} programs or groups of
programs are out of the scope of this SoK: \eg isolation of applications on a
commodity OS, whole-application sandboxing~\cite{Dunlap2022, SECCOMP, Gong1997,
Backes2015}, separation between user and kernel~\cite{Chen2008, Hofmann2013},
between VMs~\cite{Barham2003} hosting separate programs, between stakeholders
(confidential computing~\cite{Russinovich2021}), between users, or
containers~\cite{Merkel2014}.  Though these isolation works share challenges
with software compartmentalization, we focus on the latter for space reasons.
Other works such as Shu~\cite{Shu2016} complement this SoK with a broader
scope.

\vspace{-0.1cm}
\subsection{Key Differentiators} \label{overarching-characteristics}
\vspace{-0.1cm}

Three aspects are key to characterizing compartmentalizations: the \emph{choice
of subjects}, of \emph{security properties}, and of \emph{trust/threat models}.
We detail them next in our model.

\vspace{-0.1cm}
\paragraph{Subject Selection}
\label{code-data-classification}
The choice of \hypertarget{smodel}{subjects} (\S\ref{def-subject}) can be
approached in three ways: \emph{code-centric}, \emph{data-centric}, and
\emph{hybrid}.  In \ul{code-centric} (or \emph{spatial}) approaches, subjects
are defined as program instructions, and protection domains constitute code
regions.  For instance, the \emph{libjpeg} image processing library can be put
in its own protection domain~\cite{Narayan2020}.  In \ul{data-centric} (or
\emph{temporal}, \emph{horizontal}~\cite{Watson2015}) approaches, subjects are
defined as temporal units of execution, e.g., a thread or a process.
Protection domains may contain one or more of these subjects.  For example,
worker processes in a modern web server (see \S\ref{webserver-sep}) constitute
data-centric domains, all executing the same packet-processing loop in
isolation. These two strategies are not mutually exclusive: \ul{hybrid} (or
\emph{object-oriented}~\cite{Gudka2015}) approaches consider data-centric
subjects bounded within code regions, \eg a thread bounded to a specific
library.  Less popular than code- and data-centric approaches, hybrid
approaches have been explored for multi-instance code-centric
domains~\cite{Mao2011}.

\insight{
\emph{Code-centric, data-centric, and hybrid subject selections suit different
programs}.  Code-centric is appropriate when distrust can be directed at a
particular code unit (e.g., third-party code), or when secrets are associated
to specific data structures (e.g., secret keys).  Data-centric is appropriate
for programs handling mutually distrusting information flows, particularly with
per-flow secrets, e.g., a web server serving mutually-distrusting clients.
Hybrid approaches can accommodate mixed characteristics.
}

\vspace{-0.1cm}
\paragraph{Target Properties}
\label{property-based-classification}
Compartmentalization can enforce various \hypertarget{props}{properties},
including \emph{integrity}, \emph{confidentiality}, and \emph{availability}. In
this context, \ul{Integrity} guarantees that a subject cannot write out of its
protection domain.  \ul{Confidentiality} guarantees that a subject cannot read
out of its protection domain. \ul{Availability} guarantees that a subject
cannot prevent other protection domains from executing normally. Integrity is a
prerequisite for all other properties: availability generally cannot be
provided without integrity, and similar issues arise when enforcing
confidentiality without integrity~\cite{Wilke2020}.  Compartmentalization
approaches can also enforce additional properties, typically to raise the bar
against cross-compartment attacks. Among them,
\emph{\hypertarget{cccfi}{Cross-Compartment Control-Flow
Integrity}}~\cite{Sartakov2021, Lefeuvre2022} (CC-CFI) enforces valid
control-flow across compartments: cross-compartment call sites can only call
compartment entry-points they would normally call according to the global
Control-Flow Graph (CFG).  \emph{Runtime
re-compartmentalization}~\cite{Parmer2007, Nikolaev2020} enables the policy to
change at runtime to achieve more suitable security or performance trade-offs,
e.g., as the load evolves.

\vspace{-0.1cm}
\paragraph{Trust \& Threat Model(s)}
\label{def:tms}

Compartmentalization can materialize different \hypertarget{tmodel}{trust
models} by assigning subjects to domains, deciding for which domain to
prioritize least privilege, and which properties to enforce.  We observe that
all can be expressed as a composition of three key trust models:
\emph{sandbox}, \emph{safebox}, and \emph{mutual distrust}.

\label{syscall_sandboxes}
The \ul{sandbox} trust model reduces the privileges of an untrusted subject
$s_u$ to protect the rest of the system.\footnote{\emph{Sandbox} can
also refer to whole-program privilege reduction~\cite{Dunlap2022}, \eg with
seccomp~\cite{SECCOMP}, or language-based techniques~\cite{Gong1997}.  This
matches our sandbox model, but not our definition of compartmentalization
(see scope).} Least-privilege is enforced on $s_u$ only. A popular
use-case is protecting against vulnerable code, such as the request parser in a
web server. Its inverse, the \ul{safebox} model (or
\emph{vault}~\cite{Schrammel2020}), reduces the privileges of the whole system
to protect a trusted subject $s_t$.  Here, least privilege is applied on
everything but $s_t$. A typical use-case is the protection of cryptographic
secrets~\cite{VahldiekOberwagner2019}.  Finally, the \ul{mutual distrust} model
aims, for two disjoint sets of subjects $s_1$ and $s_2$, to eliminate the
privileges of both parties on the other. Here least privilege is enforced on
both sets. A typical use-case is distrust among sandboxes to increase fault
isolation~\cite{Biggs2018}.

All compartmentalization approaches rely on a \emph{Trusted Computing Base}
(TCB)~\cite{Rushby1981} to enforce compartmentalization. The TCB varies across
approaches; typically included are the CPU package and a compartmentalization
runtime (also called \emph{reference monitor}~\cite{Anderson1972}), but a
compiler, OS kernel, or additional libraries, may also be included.

Compartmentalization is applied to wide range of threats, including isolating
and recovering from adversarial and non-adversarial faults, thwarting
supply-chain attacks, or protecting other security mechanisms such as shadow
stacks.  There is thus no one single threat model of compartmentalization.

\insight{
Because compartmentalization approaches define subjects, properties, trust
models, and TCBs differently, to thwart a many different threats, \emph{there
is no one unified ``threat model of compartmentalization''.}
}

%% file: 03-approaches.tex
\vspace{-0.1cm}
\section{A Taxonomy of Compartmentalization}\label{sec:stateoftheart}
\vspace{-0.1cm}

\newcommand{\s}[1][]{\textsuperscript{\ding{93}}}
\newcommand{\sdown}[1][]{\ding{93}}

We propose to view compartmentalizing software as the combination of three key
problems ({\small{P}}):

\begin{enumerate}[noitemsep] \label{three-ps}
\item[\small{(P1)}] How to determine the right policy to enforce? Addressed by \textit{policy definition methods}.
\item[\small{(P2)}] How to express the notion of compartmentalization policies in software, programming models, and idioms? Addressed by \textit{compartmentalization abstractions}.
\item[\small{(P3)}] How to enforce policies at runtime? Addressed by \textit{compartmentalization mechanisms}.
\end{enumerate}

Nearly all works from the literature target a subset of these challenges: \eg
SOAAP~\cite{Gudka2015} addresses (\pone), SMVs~\cite{Hsu2016} (\ptwo),
Donky~\cite{Schrammel2020} (\pthree).  In fact, \ponetothree{} are rarely
addressed all at once for scale reasons.  We ground our systematization on this
division of challenges by comprehensively systematizing each challenge
(\S\ref{subsec:p1}~-~\S\ref{subsec:taxonomy-p3}).

\customparheader{Selection Methodology.} We manually filter the program of
top-tier security and systems conferences for 2003-2023 (\totalpapers papers)
through titles to obtain a list of potentially relevant works
(\potentiallyrelated papers). We choose 2003 as the point when application
compartmentalization gained visibility in research with the seminal work of
Provos~\cite{Provos2003} and Kilpatrick~\cite{Kilpatrick2003}. We then inspect
abstracts to reduce this list to \systematicsearch relevant papers.  To cover
works prior to 2003 and from other venues, we apply
\emph{snowballing}~\cite{Wohlin14} to each paper of our set to identify
\recsearch{} further works, and complete the list with \ownexpertise more from
our knowledge of the industry and literature, totaling \totalinstudy works.
\intables of them, featured in
\Cref{fig:taxonomy-algos,fig:abstraction-taxonomy,fig:taxonomy-mechanisms},
address at least one of \ponetothree{}.  Our goal is not exhaustivity (this is
not a survey), but to capture a representative set of influential
compartmentalization works. We define our data points by extracting, for
\ponetothree, key characteristics to differentiate these solutions.  This
results in 7 differentiators for \pone, 8 for \ptwo, and 7 for \pthree
(\Cref{fig:taxonomy-algos,fig:abstraction-taxonomy,fig:taxonomy-mechanisms}).
Works spanning several categories (e.g., \ponetotwo), are studied independently
from both perspectives.  Interested readers can read more about our methodology
in Appendix \hyperlink{taxonomyappendix}{A1}.

\subsection{Policy Definition Methods (P1)}
\label{subsec:p1}

\definition{
A \emph{Policy Definition Method} (PDM) is a method to define a program privilege
separation policy. PDMs identify subjects, objects, and permissions to
enforce, and may be applied to existing and new codebases.
}

In this section we systematize PDMs on the basis of our taxonomy.  It is
designed to be read along with \Cref{fig:taxonomy-algos}.

%\paragraph{Goals}
%\label{subsec:separation-goals}
%
%Applied to policy definition methods, the properties from
%\S\ref{subsec:compartmentalization-goals} translate to the following:
%
%\begin{itemize}[noitemsep]
%\item[\BC{S}] \emph{Minimizing privileges.}
%	Choosing boundaries to minimize the privilege of each
%	protection domain; Accurately identifying the privileges needed by
%	each protection domain to avoid over-privileged compartments.
%\item[\BC{S}] \emph{Maximizing interface safety.}
%	Choosing boundaries to maximize interface safety: minimizing
%	leakage, maximizing the sanitization potential of data and control flow,
%	and the ability to enforce interface semantics.
%\item[\BC{P}] \emph{Meeting performance goals.}
%	Choosing boundaries to minimize the performance cost of separation:
%	avoiding to cut critical paths, minimizing data copies, etc.
%\item[\BC{C}] \emph{Minimizing developer effort.}
%	Minimizing the expertise and effort needed from developers to
%	separate software.
%\item[\BC{U}] \emph{Maximizing completeness/soundness.}
%	Soundly identifying protection domain privileges
%	to avoid availability issues caused
%	by under-privileged compartments.
%\end{itemize}
%
%PDMs must strike a balance between these goals. Research problems lie in these
%trade-offs, which we discuss next.

\input{03-TABLE-P1}

\paragraph{Automation}
\label{automation-tradeoffs}

Automation is a key research problem in compartmentalization. It is structural
for all other characteristics discussed in this section. Through automation,
works seek to better restrict privileges or safeguard interfaces, and limit
developer effort and performance overheads.  We propose a taxonomy of four
degrees of automation:

\customparheader{\hypertarget{manual}{Manual methods}} (\manual{} in
\Cref{fig:taxonomy-algos}) rely on the expert knowledge of developers to
separate software.  Developers must define a policy at the lowest level:
\emph{which} subject is given \emph{what} permissions for \emph{what} object.
When determining permissions, manual approaches can be accurate, but are prone
to human error, resulting in \emph{false positives} (under-privileged
compartments, hurting reliability) and \emph{false negatives} (over-privileged
compartments, weakening security properties).  Similarly, performance and
interface safety widely depend on expert knowledge and human
error~\cite{Lefeuvre2022, Gudka2015}.  Overall, it is not possible to guarantee
quality or correctness with manual separation because of its reliance on human
expertise and engineering effort.  Still, much of the literature falls in the
manual category (\S\ref{p2-taxonomy}), and manually-separated software can
achieve robustness and reliability (\S\ref{sec:deployed}).

\customparheader{\hypertarget{guidedm}{Guided manual methods}}
(\lowautomation{}) such as SOAAP~\cite{Gudka2015} are manual, but provide
developers with tools to make the separation less tedious and error-prone.
Often featuring a feedback loop~\cite{Narayan2020, Gudka2015}, they guide users
to define and protect boundaries.  Guided methods can bring firm guarantees to
manually-separated software, \eg eliminating classes of confused
deputies~\cite{Narayan2020} or information leaks~\cite{Gudka2015}.

\customparheader{\hypertarget{policyr}{Policy-refinement methods}}
(\mediumautomation{}) automatically separate provided a high-level policy from
the developer.  Such policies, written in a \emph{policy language}, provide the
PDM with semantic information (e.g., annotations associating objects with a
confidentiality level~\cite{Zdancewic2002}) and/or high-level instructions
(pinpoint a library to sandbox~\cite{Bauer2021}).  Policy-refinement approaches
automatically refine this information into \emph{concrete, low-level} policies,
greatly limiting the amount of expertise required from developers, but still
requiring enough skills to provide a high-level policy.

\customparheader{\hypertarget{fullauto}{Fully automated methods}} (\fullauto{})
automatically separate software without any policy input from the user.
Instead of relying on semantic information on the software, which is hard to
infer automatically~\cite{Yun2016, Carr2017}, fully automated approaches
analyze the software for data dependencies, and apply least privilege on this
basis~\cite{Roessler2021, Wu2013}.  One drawback of relying on data
dependencies only is that, due to the architecture of software, data-dependency
relationships may exist between confidential data and untrusted parts,
resulting in partitionings that feature weaker security properties~\cite{Roessler2021}.

\insight{
The ability of PDMs to prevent confused deputies and information leaks is
directly linked to their understanding of software semantics, which diminishes
as automation increases.  Thus, \emph{policy refinement (\mediumautomation{})
and full automation (\fullauto{}) trade off security for developer effort.}
}

% NOTE: regarding "resulting in partitionings that feature weaker
% properties...", we cite uSCOPE without a lot of explanations. I'd love to add
% more but we don't have space for now. I'm specifically referring to Section
% 8.8, which recommends manual tweaking of object weights to limit this effect
% of "data-dependencies between confidential data and untrusted components".

\paragraph{Policy Languages}
%      -> used at least in Provos 2003a with the same meaning

Policy languages allow developers to describe high-level policies to
\hyperlink{guidedm}{Guided manual} (\lowautomation{}) or
\hyperlink{policyr}{Policy-refinement} (\mediumautomation{}) PDMs. They are key
to achieve intermediate degrees of automation, providing PDMs with an
understanding of program semantics and trust relationships which is otherwise
hard to extract automatically~\cite{Yun2016}. We distinguish between two types
of policy languages: \emph{annotations}, and \emph{placement rules}.
\ul{Annotation}-based policies provide fine-grain semantics on subjects and
objects, such as describing shared, confidential, or sensitive
entities~\cite{Zdancewic2002, Brumley2004, Gudka2015, Liu2015, Liu2017,
Liu2019a, Lefeuvre2022} (e.g., object \customtt{key} is confidential, function
\customtt{parse()} is sensitive), past vulnerabilities~\cite{Gudka2015}, or
performance goals and bottlenecks~\cite{Gudka2015, Liu2019a}. Annotations are
tightly coupled with program code such that they may be provided by external
dependency vendors~\cite{Gudka2015}. Conversely, \ul{placement rule} languages
provide coarse-grain, high-level descriptions of component trust
relationships~\cite{Zdancewic2002, Bauer2021, Lefeuvre2022} and/or building
rules~\cite{Liu2019a, Clements2018} (e.g., place libraries $X$ and $Y$ in
separate domains).  Placement rules are independent from program code, and are
thus more commonly provided by system integrators.  They can be expressed in
many ways, including human-readable JSON files~\cite{Bauer2021, Lefeuvre2022},
or integrated into the build-system~\cite{Huang2022}. Both classes are not
mutually exclusive and may be combined by PDMs~\cite{Liu2019a, Lefeuvre2022}.

\insight{
Annotations and placement rules have \emph{different expressivity}: annotations
express local, low-level semantics, whereas placement rules express full-system
properties.  Both suit \emph{different threat models}: since annotations are
tightly coupled with program code, they may be vulnerable to supply-chain
attacks, unlike placement rules. This provides strong incentives to combine the
two approaches, which is not a common practice (\cf \Cref{fig:taxonomy-algos}).
}

\paragraph{Separation Granularities}
\label{granularity-tradeoffs}

Many granularities have been explored (\Cref{fig:taxonomy-algos}):
functions~\cite{Wanninger2022}, libraries~\cite{Bauer2021}, linkage
units~\cite{Almatary2022}, drivers~\cite{Huang2022}, software
packages~\cite{Ghosn2021}, etc. These choices are guided by design decisions.
On the one hand, \ul{finer granularities} of separation make it possible to
better enforce least privilege, or reach boundaries more favorable to
performance.  Fine granularities may also be necessary to tackle certain
vulnerabilities, e.g., Heartbleed~\cite{Durumeric2014}.  Conversely, for some
separation approaches, the threat model itself may be defined at a \ul{coarse
granularity}; larger components such as libraries or packages are valid units
of trust in real-world scenarios such as supply-chain
attacks~\cite{Vasilakis2018}. Operating at coarser granularities may also
reduce developer effort and expertise requirements: as granularities become
finer, it becomes more complex to express policies and reason about them;
higher-level boundaries such as libraries are more intuitive separation units
than arbitrary internal functions. Finer granularities may also negatively
impact interface safety: as boundaries are set at internal, less encapsulated
software layers~\cite{Roessler2021}, interfaces are more exposed to confused
deputies and harder to safeguard~\cite{Lefeuvre2022}. Finally, finer
granularities pose technical challenges of state explosion~\cite{Lefeuvre2022},
performance~\cite{Huang2021} and analysis/clustering~\cite{Roessler2021}.

\paragraph{Analysis Techniques}
\label{analysis-techniques}

Automated PDMs ({\footnotesize
\lowautomation{}\hspace{0.05cm}\mediumautomation{}\hspace{0.05cm}\fullauto{}})
employ a range of static, dynamic, and hybrid techniques which, through subtle
trade-offs, strongly impact final properties.

\customparheader{Static methods.} When determining permissions, approaches
based on static analysis are \emph{complete} but conservative: separated
software is guaranteed to function correctly, but compartments may be left
over-privileged due to the fundamental imprecision of static
analysis~\cite{Liu2019a, Clements2018}. When applied to performance analysis,
static approaches can provide useful~\cite{Liu2019a} albeit imprecise
performance metrics~\cite{Gudka2015}. Applied to improve interface safety,
static analysis approaches can detect potential issues at scale and
systematically~\cite{Narayan2020}, but cannot yield precise impact
metrics~\cite{Hu2015}.

\insight{
Though the problem of over-privilege in static analysis is well-known and its
impact characterized~\cite{Lefeuvre2023}, it has not been \emph{quantified},
such that it is unclear to what extent static PDMs trade security for
developer effort.
}

\customparheader{Dynamic methods.} When determining permissions, dynamic
methods guarantee that permissions granted to compartments are strictly
necessary. However they are \emph{incomplete}: due to their fundamental
coverage problem~\cite{Roessler2021, Liu2019a} domains may be left
under-privileged, so that separated software may not function correctly anymore
for all workloads.  Similarly, dynamic methods can provide precise performance
estimates~\cite{Lefeuvre2022, Roessler2021} but only on covered workloads.
Applied to interface safety, they can detect vulnerabilities and their concrete
impact, but not systematically~\cite{Lefeuvre2023}.  Due to the problem of
incompleteness, very few PDMs are dynamic (5 out of 27 in
\Cref{fig:taxonomy-algos}).

\customparheader{Hybrid methods.} Static and dynamic methods do not compose
well, as it is hard to utilize the delta between static and dynamic results.
On the privilege detection side, some use this delta as the \emph{suspicious
subset}~\cite{Liu2015}, authorizing but reporting uses, which poses usability
questions.  Others use dynamic results to optimize
compartmentalizations~\cite{Brumley2004}, to predict performance accurately
with an otherwise static approach~\cite{Gudka2015}, or to measure information
flow~\cite{Liu2019a}.

\paragraph{Subject Selections} \label{no-data-centric-sep}
Policies can apply to different types of subjects: code-centric, data-centric,
hybrid, each suiting different software characteristics (see
\S\ref{code-data-classification}). This choice strongly impacts the design of
PDMs, particularly when targeting automation. Surprisingly all automated PDMs
are code-centric, and non-code-centric PDMs all fall into the guided- or manual
categories (\cf \Cref{fig:taxonomy-algos}).

\insight{
\emph{Research in PDMs largely focuses on code-centric separation.} We speculate that
this is caused by the lesser popularity of data-centric approaches (we repeat
this observation in \S\ref{p2-taxonomy}), but also to the greater complexity of
data-centric separation: while many automated code-centric PDMs assume
single-threaded programs, data-centric requires concurrent separation,
complexifying the analysis~\cite{Liu2017}. This calls for more research in
automated PDMs for data-centric and hybrid subjects.
}

\paragraph{Genericity}

Most PDMs specialize on classes of programming languages (\cf
\Cref{fig:taxonomy-algos}), for several reasons. First, some focus on
domain-specific problems or threat models: \eg pointer aliasing in
C~\cite{Liu2017}, untrusted packages in modern languages~\cite{Ghosn2021}.
Second, specializing on language classes allows PDMs to leverage language
specificities: their type system to detect API sanitization
needs~\cite{Narayan2020}; their memory safety~\cite{Zdancewic2002, Chong2007}
or interpreted nature~\cite{Vasilakis2018, Ghosn2021} to simplify boundary
detection; or the overall system architecture~\cite{Huang2017} to make
assumptions on boundaries. There is, for instance, a vast body of work (only
partially covered in \Cref{fig:taxonomy-algos} for space reasons) specifically
targeting 3rd-party Android library code~\cite{Sun2014, Seo2016, Hu2021}, with
some~\cite{Pearce2012, Shekhar2012, Zhang2013, Liu2015a, Huang2017}
specializing entirely on advertisement libraries.  This is well-covered by Acar
et al.~\cite{Acar2016}.

\subsection{Compartmentalization Abstractions (P2)}
\label{p2-taxonomy}

\input{03-TABLE-P2}

\definition{
A \emph{compartmentalization abstraction} defines and implements primitives to
express separation policies in a program. Depending on the semantics of these
primitives, abstractions may be used to express different types of subjects, trust models,
properties, etc.
}

We first contribute a model to characterize the core primitives of
compartmentalization abstractions. We then leverage this model to systematize
existing approaches. This section is designed to be read along with
\Cref{fig:abstraction-taxonomy}.

\paragraph{A Model of Compartmentalization Abstractions}
\label{p2-primitives}

Compartmentalization abstractions instantiate the notion of a
\emph{compartment}, defining the type of subjects compartments may isolate,
properties they may enforce, and the trust models that may be implemented.
They must also define five primitives: \emph{\customtt{CREATE}} and
\emph{\customtt{DESTROY}} a compartment (defining the semantics of compartment
lifetime management, the default permissions of new compartments, among
others); \emph{\customtt{CALL}} and \emph{\customtt{RETURN}} from a compartment
(defining cross-compartment control-flow semantics); and
\emph{\customtt{ASSIGN} privileges} (granting and revoking permissions across
compartments, resource ownership).  Abstractions achieve trade-offs by
controlling the \hypertarget{semant}{semantics} of these primitives.

Not all five primitives must be exposed to developers; when they are exposed,
we refer to them as having \emph{explicit} semantics.  Inversely,
\emph{implicit} primitives are handled automatically under the hood.  Implicit
semantics are common in abstractions that are coupled with the PDM (\eg
automatic \customtt{CREATE}/\customtt{DESTROY}, transparent \customtt{CALL}s,
automatic \customtt{ASSIGN}). Additionally, abstractions may provide other
primitives, \eg to support fault tolerance.  We now detail the core primitives,
focusing on \customtt{CALL/RETURN} and \customtt{ASSIGN} for space reasons.

% having this would be nice, but do we have space...
%\customparheader{\customtt{CREATE/DESTROY}.}

\customparheader{\customtt{CALL/RETURN}.}
Regardless of their implementation, \customtt{CALL} and \customtt{RETURN} must
meet minimum safety requirements: 1) guaranteeing the validity of control-flow
entry-points in compartments (compartments should not \customtt{CALL} or
\customtt{RETURN} to arbitrary code in the context of other compartments); 2)
switching call stacks and clearing registers appropriately to avoid
unintentional leakages; but also 3) ensuring that the abstraction composes
safely with other system interfaces.

Cross-compartment control flow can be approached \emph{synchronously}, or
\emph{asynchronously}. In the \ul{synchronous} case, \customtt{CALL} and
\customtt{RETURN} are semantically similar to a local function call and thus
transparent to separated programs. In the \ul{asynchronous} case,
\customtt{CALL} and \customtt{RETURN} abandon function-call semantics: the
execution of the caller domain continues after the call, and the return is
processed by the caller similarly to a separate message~\cite{Lampson1974}
(e.g., as part of an event loop).  Asynchronous semantics are less popular in
\Cref{fig:abstraction-taxonomy}.  This is likely due to the fact that they are
more disruptive compatibility-wise: applications must be ``structurally'' aware
of the separation, and redesigning for asynchronous behavior is
non-trivial~\cite{Watson2015}.  Still, asynchronous semantics can be beneficial
to performance, as their non-blocking nature can mask boundary-crossing
delays~\cite{Soares2010}.  For certain target properties such as availability,
\customtt{CALL} and \customtt{RETURN} ultimately need distantiating from
function call semantics, as new error types appear: timeout, callee compartment
failure, etc. These translate into asynchronous features in call semantics that
may otherwise be synchronous~\cite{Almatary2022}.

\customparheader{\customtt{ASSIGN}.} \customtt{ASSIGN} semantics are structured
by two fundamental approaches to communicating data~\cite{Attiya1995}:
\emph{message passing}, and \emph{shared data} (or \emph{message/object}
systems~\cite{Lampson1974}). With \ul{message passing}, domains share data
across boundaries via messages over a communication channel (e.g., POSIX
sockets or pipes).  This means that objects are not only systematically copied,
but also marshalled, and as part of this, potentially serialized and checked.
This makes message-based solutions rather disruptive compatibility- and
performance-wise: they do not map to natural shared-memory semantics found in
applications, and require copies.  Still, systematic copying and checking
greatly benefits security~\cite{Lefeuvre2023}.  With \ul{shared memory},
protection domains both have privileges over shared memory, and communicate via
loads/stores.  Although copies can still be made systematic by the
abstraction~\cite{Narayan2020}, it is not the norm: copies are costly, it is
thus enticing to avoid them whenever possible.  This makes shared memory much
less disruptive compatibility- and performance-wise, but potentially deceptive
security-wise~\cite{Lefeuvre2023}. Note that we describe exposed abstraction
semantics here; under the hood shared-data can be implemented on top of message
passing, and vice-versa~\cite{Attiya1995}.

\insight{
Whereas compartmentalization abstraction semantics were historically
asynchronous and message-passing based, new trends in retrofitting
separation shifted the focus to synchronous and shared-memory
approaches. \emph{This comes at a non-trivial security and performance cost.}
}

\paragraph{Trust Models}
\label{p2-tms}

Safebox, sandbox, and mutual distrust (\S\ref{def:tms}) are all represented in
\Cref{fig:abstraction-taxonomy}.  \ul{Sandbox} and \ul{mutual distrust}
abstractions all support scenarios with an arbitrary number of compartments.
Although Shreds~\cite{Chen2016} supports arbitrary \ul{safebox} scenarios, all
other safebox abstractions are limited to two compartments (trusted \vs
untrusted, ``\emph{Dual World}'' in \Cref{fig:abstraction-taxonomy}). This
shows a lesser interest in applying distrust among trusted entities.  Though
safebox and sandbox abstractions can both be implemented on top of arbitrary
mutual distrust (and can thus be seen as special cases), the presence of
safeboxes or sandboxes only, or of a fixed number of compartments, considerably
simplifies their semantics.  Each presents trade-offs. Mutual distrust
abstractions can express true least privilege.  However, this comes at a
performance cost, as they must enforce integrity and other properties on all
compartments, whereas one-sided distrust models (safebox, sandbox) must only
enforce them for the trusted side.  Mutual distrust also makes interface
hardening generally more challenging and costly in
performance~\cite{Lefeuvre2023}.

\paragraph{Enforcing More or Fewer Properties}
\label{subsec:abstractions-properties}

All abstractions in \Cref{fig:abstraction-taxonomy} enforce integrity, a
prerequisite for other properties (\S\ref{property-based-classification}).
Most target confidentiality, and some target availability, runtime
re-compartmentalization, or interface safety.

\customparheader{Confidentiality.} Since nearly all abstractions provide
confidentiality, we focus on the impact of \emph{not} doing so. Not providing
confidentiality does not cause structural changes in abstractions~\cite{MONZA}.
It may benefit performance, as it enables zero-cost read-only data
sharing, i.e., fewer copies at compartment boundaries.  It also simplifies
separation: only write-shared data require explicit sharing.  On the downside,
not providing confidentiality vastly reduces the abstraction's ability
to counter information leakages: the only remaining barrier is the
compartments' ability to limit data exfiltration vectors. It may also be
detrimental from an interface-safety viewpoint, defeating randomization
(thus easing cross-compartment exploits). Lastly, avoiding copies at
boundaries may make the system more prone to shared-memory
TOCTOU~\cite{Lefeuvre2023}. Overall, except for legitimate integrity-only
use-cases (e.g., shadow-stack protection~\cite{Koning2017}), giving up
confidentiality trades security for performance and compatibility.

\customparheader{Availability.} \label{p2-avail} Abstractions may go beyond
fault isolation and target cross-compartment fault tolerance. This brings many
well-known challenges from the fault-tolerance and distributed systems
fields~\cite{VanSteen2020}: \eg avoiding, detecting, recovering from resource
exhaustion, and from various failures (omission, Byzantine).
Particularly relevant to compartmentalization are resource ownership
problems~\cite{Narayanan2020a} (when restarting a domain, can shared resources
be released?), and state coherence issues~\cite{Boos2017a} (the state of
restarted domains may be incoherent with that of other components), as
compartmentalized components, particularly when retrofitting, are typically
less encapsulated~\cite{Lefeuvre2023}. Overall, availability differs from
properties such as integrity and confidentiality in that it is a property of
the whole system, not a compartment-local property: to achieve availability
with compartmentalization, the interaction (and failure) of all domains
must be considered at once, instead of separately.

To tackle these challenges, abstractions take a vast variety of approaches,
whose exhaustive listing outreaches the scope of this paper: bounded execution
in resources~\cite{Nikolaev2013} or time~\cite{Gudka2015, Almatary2022}, and
generally performance isolation~\cite{Gupta2006} to tackle resource exhaustion;
leveraging type systems for resource ownership~\cite{Narayanan2020a}; proposing
manually-designed interface wrappers~\cite{Swift2006, Narayanan2020a},
per-component fault-handlers~\cite{Almatary2022}, careful TCB and interface designs
to store state outside of domains~\cite{Nikolaev2020}, or recursive restarting
of relevant components~\cite{Nikolaev2013} for state coherence.  Because the problem
is hard, all require expert understanding of fault domains, rely on a
variable amount of manual engineering, and not all approaches are complete;
e.g., Google SAPI~\cite{GOOGLE_SAPI} automatically re-iterates failed
calls and restarts faulty components, but does not consider state coherence
problems.

\insight{
\emph{Very few compartmentalization abstractions target availability}.  We
speculate that this is due to the additional complexity of fault tolerance, as
discussed above. This is problematic: as we show in \S\ref{sec:deployed}
availability is a common need of real-world deployments.  \emph{This calls for
more work on fault-tolerant compartment abstractions.}
}

\customparheader{Runtime Re-Compartmentalization.} Some abstractions also
support changing the policy enforced at \emph{runtime}~\cite{Parmer2007,
Parmer2012, Nikolaev2020}, \eg to adapt policies to evolving requirements in
performance and fault isolation.  Beyond technical challenges of achieving
transparent, fast re-compartmentalization, we observe that this poses
non-trivial security challenges in adversarial contexts: assuming attackers can
wait for the weakest fault-isolation profile to be instantiated, or influence
workloads to trigger such profiles (e.g., generate more, or different network
traffic), then the overall security properties are that of the \emph{weakest}
profile achievable at any point in time.  This may be true even when attackers
cannot wait; since component states are preserved across policy changes, any
undetected corruption triggered by an attacker during a ``strong'' profile will
eventually reach other components when weaker profiles trigger, similar to
delay attacks~\cite{Yagemann2021}. This limits applications of
re-compartmentalization to non-adversarial scenarios.

\customparheader{\hypertarget{abstractionisafety}{Compartment-Interface Hardening}.}\label{par:abstractionisafety}
Although \hyperlink{isafety}{interface safety} is a key problem of
compartmentalization, most abstractions consider it orthogonal to their mission
(\Cref{fig:abstraction-taxonomy}). Works often transfer the responsibility to
PDMs by assuming a safe separation policy, or to downstream developers by
assuming hardened components.  Yet, though the purpose of compartmentalization
abstractions is not to help users defining security policies (this is the role
of PDMs), they can contribute to interface safety by \emph{making it harder to
implement unsafe policies}, thus ensuring that compartment interfaces are free
of certain classes of confused deputies.  They may do so at various levels, \eg
by \emph{enforcing restrictions on interface definitions}, such as restricting
interface-crossing types~\cite{Mao2011, Narayan2020} or enforcing points-to
ranges for interface-crossing pointers~\cite{Mao2011, Lind2017, Narayan2020,
Khandaker2020, Lefeuvre2023a}; by \emph{forcing the presence of checks on interface-crossing
data}, forcing users to write checks~\cite{Narayan2020}, or checking
automatically when possible~\cite{Mao2011, Lind2017, Narayan2020}; or by
\emph{enforcing restrictions on cross-compartment control-flow}, providing
primitives to specify and enforce API call ordering~\cite{Narayan2020}, and
enforcing additional properties such as \hyperlink{cccfi}{CC-CFI} to raise the bar
for cross-compartment attacks~\cite{Sartakov2021, Lefeuvre2022}.  These
measures are not orthogonal to the core mission of compartmentalization
abstractions, as they generally cannot be implemented independently: without
knowledge of compartments mappings, it may be impossible to verify
pointers and indexes; to check reference types or call ordering; to implement
\hyperlink{cccfi}{CC-CFI}; etc. Lefeuvre~\cite{Lefeuvre2023} and
Hu~\cite{Hu2015} provide deeper coverage of the topic of interface safety in
compartmentalization.

\insight{
\hypertarget{insightabstractioncivs}{
There is a widespread misconception that interface safety is orthogonal to the
mission of compartmentalization abstractions.  \emph{More work is needed on
abstractions that contribute to compartment-interface hardening}.
}
}

\paragraph{Composing \ponetothreeverbose}
\label{abstraction-specialization}

\hfill \break
\vspace{-0.2cm}
\hfill \break
\customparheader{Subject Selection.} Similarly to PDMs, most abstractions
specialize towards certain subject types (\Cref{fig:abstraction-taxonomy}).
Abstractions focusing on \ul{code-centric} models assume a direct mapping
between code and compartment. As a result, \customtt{CREATE} and
\customtt{DESTROY} are implicit: developers are not provided with explicit
controls to manage compartment lifetime, and the abstraction is \emph{static},
\ie the number and content of compartments is known at compile time.
Conversely, those focusing on \ul{data-centric} models feature explicit
\customtt{CREATE} and \customtt{DESTROY}, and are \emph{dynamic}, \ie the
number and content of compartments may not be known at compile time and depend
on the control flow taken at runtime.  Abstractions that are \ul{hybrid} are
similar to data-centric abstractions but allow restricting the code available
to a compartment; they can be used to implement both code- and data-centric
separations.

\customparheader{\hypertarget{granty}{Granularities}.} Abstractions may also
specialize towards specific domain granularities. These decisions are embodied
in \customtt{CREATE} and \customtt{CALL} semantics which define the granularity
at which compartments may be created and entered.  This specialization comes
with various goals: target properties may imply a granularity (\eg fault
resilience implies coarser grains), threat models may imply a granularity
(library sandboxing $\rightarrow$ library granularity), or abstractions may be
coupled together with a PDM that itself restricts granularity.

\customparheader{Mechanisms.} Many abstractions are tightly coupled with a
specific mechanism (\Cref{fig:abstraction-taxonomy}).  This is typically done
to better leverage mechanism-specific properties such as safe copy-less sharing
for capabilities, or strong typing, points-to knowledge, and provable
termination for safe languages. Tightly coupling with mechanisms has drawbacks:
beyond making abstractions unusable without their related mechanism, this
creates a mechanism dependency in downstream programs, which curbs efforts to
roll out new mechanisms (\eg{} \customtt{fork()} makes it hard to replace the
page table~\cite{Baumann2019}). These limitations incited a recent trend
towards mechanism-agnostic abstractions~\cite{Koning2017, Narayan2020, Lefeuvre2022}.

\insight{
\hypertarget{holisticinsight}{Design decisions} on either of
\ponetothreeverbose have implications across the stack. The split of
\ponetothree has its limitations and \emph{all three problems must eventually
be considered together to achieve harmonious solutions.}
}

\paragraph{Composing with Other Abstractions}
\label{subsec:composing}

\hfill \break
\vspace{-0.2cm}
\hfill \break
\indent\customparheader{Threads.}
Compartmentalization abstractions must define a threading model for
compartments. We classify threading models as either \emph{orthogonal} or
\emph{coupled}.  In the \ul{orthogonal} case~\cite{Mao2011,
VahldiekOberwagner2019, Lefeuvre2022}, threads cross protection domains as they
\customtt{CALL} and \customtt{RETURN}.  To ensure safety, these operations must
guarantee that the underlying thread state is updated to reflect the crossing,
and carefully define the behavior of thread local storage.  By definition,
orthogonal threading models exclusively suit \emph{code-centric} approaches.
In the \ul{coupled} case~\cite{Bittau2008, Wang2015, Hsu2016}, threads are
immutably assigned to a protection domain at their creation, and
\customtt{CALL} spawns a new thread in the desired protection domain. Coupled
threading models suit any subject selection.

\customparheader{CPU Privilege Levels (Rings).}
In \ul{userland}, semantics of compartmentalization abstractions are heavily
influenced by the presence of the user/kernel interface. Compartmentalization
abstractions may~\cite{Hsu2016} or may not~\cite{VahldiekOberwagner2019} be
exposed as part of the user/kernel boundary for performance, security, or
compatibility reasons. As we discuss \hyperlink{compositionsscall}{next},
user-mode abstractions must harmoniously compose with kernel interfaces such as
processes, signals, or system calls. Different factors shape \ul{kernel} and
\ul{hypervisor} mode compartmentalization abstractions. Kernel and hypervisor
codebases are often designed assuming ambient privilege on hardware and user
data, making the TCB (and retrofitting) less obvious than in userland. The need
to integrate with low-level events such as interrupts brings even more specific
challenges that make it necessary for abstractions to integrate deeper in
kernel and hypervisor designs~\cite{Dautenhahn2015, Narayanan2020,
Lefeuvre2022}, encompassing boot, scheduling, memory management, or interrupt
handling.

\customparheader{Processes.}
Though processes are themselves a compartmentalization abstraction, they are
also used by programs for reasons other than protection~\cite{Baumann2019}.
Several prior works showed that composing compartmentalization abstractions
with processes is error-prone~\cite{Connor2020, Schrammel2022}.  User-mode
compartmentalization abstractions must thus take special care defining how they
compose with processes.  To ensure safety, approaches may intercept and forbid
attempts to spawn new processes~\cite{Schrammel2022}, at the expense of
application compatibility.

\customparheader{\hypertarget{compositionsscall}{System Calls \& Other System
Interfaces.}} \label{p2-syscalls}
We discussed earlier (Insight \hyperlink{insightabstractioncivs}{9}) how compartmentalization
abstractions can contribute to safeguarding intra-program compartment
interfaces. Yet, interfaces with the \emph{rest of the system} are also major
\hyperlink{isafety}{interface safety} weak spots~\cite{Savage1994, Connor2020,
Voulimeneas2022, Lefeuvre2023}. These include, in user-mode, system calls,
other kernel abstractions (pseudo file-systems~\cite{PROCFS, SYSFS}, files,
sockets), but also interfaces exposed by other applications on the
system~\cite{Chen2007}.  Safeguarding these interfaces is non-trivial:
mechanisms come with different protection needs (e.g., protection keys with
``PKU pitfalls''~\cite{Connor2020}); protection needs are ABI-bound and thus
vary across OSes, configurations, and architectures; protection must be
maintained as these ABIs evolve; the protection effort itself comes with
application compatibility problems (\eg precisely detecting the OS features
required by individual application components is hard~\cite{DeMarinis2020,
Canella2021}); and this protection often results in noticeable performance
overheads~\cite{Connor2020}.  Compartmentalization abstractions either attempt
to solve this problem through careful composition with the
kernel~\cite{Gudka2015, Narayan2020, Litton2016, Ghosn2021, Schrammel2022,
Peng2023}, or scope it out as a separate problem~\cite{Liu2017,
VahldiekOberwagner2019, Hedayati2019} addressed by other
solutions~\cite{Provos2003a, Watson2010, SECCOMP, PLEDGE}.  Still, there is a
growing consensus that user-mode compartmentalization abstractions should be
designed hand in hand with kernel and broader system interfaces to maximize
interface safety~\cite{Peng2023, Schrammel2022, Connor2020}.

\vspace{-0.1cm}
\subsection{Compartmentalization Mechanisms (P3)}
\label{subsec:taxonomy-p3}

\input{03-TABLE-P3}

\vspace{-0.05cm}
\definition{
A \emph{Compartmentalization mechanism} enforces separation, as defined
by PDMs and implemented through compartmentalization abstractions, at runtime.
}

Next, we concretize this definition by modeling the fundamental primitives a
compartmentalization mechanism must provide.  Then, we show how mechanisms
approach these primitives to reach trade-offs, going through
\Cref{fig:taxonomy-mechanisms}.

\vspace{-0.05cm}
\paragraph{A Model of Compartmentalization Mechanisms}

\label{p3-conditional} At the core, a mechanism must define at least two
primitives: a \emph{protection domain} primitive, and a \emph{communication}
primitive.  The former enforces isolation across protection domains and must at
least guarantee integrity (\S\ref{property-based-classification}).  The latter
must be able to transmit bits bidirectionally across compartments, and enforce
compartment control-flow entry points.  Communication primitives can be
implemented in many ways~\cite{Lampson1974}: message passing, shared memory,
specialized control-flow operations (e.g., cross-compartment call).  Still, a
simple message passing primitive suffices for compartmentalization.
\customtt{CALL}, \customtt{RETURN}, and \customtt{ASSIGN}
(\S\ref{p2-primitives}) can all be implemented on top of it, and
\customtt{CREATE}/\customtt{DESTROY} can be left implicit. If we expand the
mechanism with a third primitive to \customtt{CREATE} domains, we obtain a
general compartmentalization mechanism; \customtt{DESTROY}, and other
mechanism-specific primitives, may also be supported.

A mechanism may not fulfill all properties necessary to be suitable for
compartmentalization, in which case we call it \emph{conditioned}.  For instance,
PKU~\cite{INTEL_MPK_SDM} is conditioned since compartment entry points cannot
be enforced in without additional measures to monitor key-editing
instructions~\cite{VahldiekOberwagner2019, Hedayati2019}. Other conditioned
cases include PT-based protection in kernel mode~\cite{Dautenhahn2015}, or
bounds-checking~\cite{INTEL_MPX, Koning2017}.  Not all mechanisms discussed
next were specifically designed for compartmentalization (\eg TEEs,
bounds-checking). Still, all are either suitable or conditioned, and have been
leveraged for compartmentalization in practice.

\insight{
Many mechanisms used for compartmentalization were not designed for that aim,
causing them to be conditioned. This \emph{has important implications on how we
evaluate their security or performance cost.}
}

\paragraph{Trust Models \& TCB}
Mechanisms themselves are designed for a given trust model (\emph{TM} in
\Cref{fig:taxonomy-mechanisms}). Here it is sufficient to distinguish between
\emph{single} distrust (covering both safebox and sandbox), and \emph{mutual}
distrust. For instance, the PT supervisor bit enforces single distrust by
protecting one subject (the kernel) from others (users), whereas the page-table
enforces mutual distrust by separating arbitrary sets of subjects. This
constrains the trust models which abstractions can implement on top of these
mechanisms (\S\ref{p2-tms}).

Mechanisms also influence the content of the TCB. In the general case
(\emph{Full} in \Cref{fig:taxonomy-mechanisms}), TCBs of compartmentalized
systems include (part of) the workload, compiler, loader, system software,
firmware, CPU package, and physical environment. TEEs exclude firmware and
physical environment from the TCB~\cite{Tsai2017, Ge2022}. From a
compartmentalization view, the TCB can be the only difference between otherwise
similar mechanisms, e.g., confidential VMs \vs EPT.

\paragraph{SW / HW} Mechanisms can be implemented in hardware (ASICs, FPGAs,
simulators) or in software. Compartmentalization advances have historically
been driven by progress in hardware, which enabled separation granularities and
security properties previously unreachable in software.  Still, hardware is not
a silver bullet: hardware takes time to reach end-devices, can be
cost-prohibitive, and, as we discuss below, tends to be heterogeneous due to
the lack of industry standard.  Thus, in recent years, software-based
approaches building on commodity hardware (MMU) have been particularly
successful in popularizing compartmentalization practices (\eg SFI with
WebAssembly~\cite{Haas2017}).

\paragraph{Permissions Enforced} \label{p3-penforced} Mechanisms enforce
different combinations of four primitive permissions: \emph{read},
\emph{write}, \emph{execute}, and \emph{address} (create pointers to). Subsets
are common: most protection key approaches~\cite{IBM_ZOS_PKEYS, INTEL_MPK_SDM,
ARM_MPU, Schrammel2020} do not protect instruction fetch; PT approaches do not
support all combinations of R/W/X~\cite{INTEL_PT_PROTECTION}; and
capabilities~\cite{Carter1994, Heiser1999} also protect addressing, which few
other mechanism classes enforce.  These specificities are the result of
trade-offs.  Security benefits from more expressiveness in permissions, and
properties such as addressing also benefit interface safety by thwarting certain
classes of confused deputies by construction~\cite{Lefeuvre2023}. On the other
hand performance is sensitive to implementation constraints which may require
to trade off security: PT entries are limited in size, and using more bits to
represent more permissions requires separate tables, degrading performance;
capabilities often trade off performance (e.g., through cache
pressure~\cite{Woodruff2014}). This poses non-trivial problems to abstractions,
which must map diverse levels of permission expressiveness to the previously
described high-level properties (\S\ref{subsec:abstractions-properties}).

\paragraph{Enforcement Granularity} Mechanisms enforce permissions at varying
memory granularities. At the extremes, enforcement may be done at the
granularity of the entire physical memory~\cite{Rushby1981}, or at byte
granularity~\cite{Watson2015}. Others (\Cref{fig:taxonomy-mechanisms}) operate
on pages, words, 128-bit chunks, etc. Granularity choices too trade off
performance, security, and compatibility. Whereas finer granularities approach
least privilege more closely, performance and memory footprint benefit from
coarser granularities due to implementation constraints: supporting finer
grains implies storing more permission information, potentially increases the
complexity of permission checks, or of protection domain instantiation.

\paragraph{Number of Domains} The domain creation primitive may restrict the
maximum number of domains: protection keys support, depending on the
implementation, a few~\cite{ARM_MPU} to thousands of~\cite{Schrammel2020}
domains. The domain creation primitive may also not exist at all: physical
separation~\cite{Rushby1981} and TrustZone~\cite{ARM_TRUSTZONE} rely on a fixed
number of physically separated domains.  Other mechanisms may not limit the
number of domains, but scalability in performance and resource usage limit it
in practice.  These decisions too are trade-offs: for protection keys, the
number of domains can be increased at a high performance cost~\cite{Park2019,
Gu2022b, Yuan2023}.

\insight{
Mechanisms feature very heterogeneous properties (such as the permissions
enforced, the enforcement granularity, or domain count). This reinforces
Insight \hyperlink{holisticinsight}{10} on the need to approach
compartmentalization holistically.
}

\paragraph{Performance} \label{subsec:mechanism-performance}
Mechanisms impact performance in many ways: latencies of compartment switching,
creation, modification, and destruction; locality and cache effects;
domain-crossing sanitization costs (which mechanisms may accelerate);
scalability (growing costs with domain count or size); and other
mechanism-specific runtime overheads (such as memory access cost or generated
code size)~\cite{Koning2017, Jangda2019}.  The performance profile of
mechanisms varies widely: PKU domain switches are unprivileged and thus fast,
but domain creation and modification requires a costly
trap~\cite{INTEL_MPK_SDM}. With software memory encryption, domain switches are
expensive (compartment encryption and decryption), but their creation and
modification is an unprivileged bookkeeping operation.

There is a focus on domain switch latencies in the literature, as they often
(yet not always) dominate compartmentalization performance
overheads~\cite{Kolosick2022}. Domain switch costs result from design decisions
leading to security, performance, and compatibility trade-offs: Is the switch a
\emph{privileged primitive}, i.e., do cross-domain switches require trampolines
or traps to the TCB with elevated privileges? Should domain switches require
trampolines at all, or should they be encoded with separate load and store
instructions~\cite{Frassetto2018, Mogosanu2018}? Should switches be made faster
at the expense of, e.g., compartment creation costs? How deeply can
compatibility (with existing compilers, OS kernels, and programs) be broken?
How generic (granularity, domain count) must the mechanism be?  For these
reasons, comparing mechanism domain-switch costs can be deceptive: conditioned
mechanisms such as PKU~\cite{INTEL_MPK_SDM} are fast but insufficient by
themselves to guarantee safety, requiring combination with additional software
techniques~\cite{Park2019, Gu2022b} which come with costs and trade-offs of
their own.

\insight{
\hypertarget{performanceinsight}{
There is a strong focus on domain-switch latencies when evaluating mechanism
performance.} Yet, this disregards many other cost factors relevant in real-world
deployments. Domain-switch cost comparisons are also commonly done between
conditioned and unconditioned mechanisms (\eg PKU \vs CHERI), which is unsound.
}

\paragraph{Side Channels} \label{subsec:side-channels}

All software-exploitable side channels are relevant to compartmentalization
threat models.

Transient execution side channels, starting with Spectre~\cite{Kocher2019} and
Meltdown~\cite{Lipp2018}, demonstrated the ability to break the confidentiality
and integrity (and thus availability) properties of mechanisms such as the page
table~\cite{Canella2019}, Intel PKU~\cite{Canella2019}, Intel
MPX~\cite{Canella2019}, ARM PAC~\cite{Ravichandran2022}, or ARM
TrustZone~\cite{Cerdeira2020}. For instance, Meltdown-PK~\cite{Canella2019}
allows attackers to fully bypass Intel PKU.  Software-based
compartmentalization mechanisms are equally vulnerable~\cite{Narayan2021,
Cauligi2022, Jin2024}, enabling attackers to bypass the confidentiality and
integrity properties of WebAssembly, eBPF, and others~\cite{Cauligi2022}.

Side channels with sequential execution semantics, such as timing side
channels~\cite{Kocher1996} or cache-timing side channels~\cite{Yarom2014}, as
well as software-exploitable power consumption side channels~\cite{Kogler2023},
can also be leveraged to bypass the confidentiality properties of
compartmentalization.

Many side-channel mitigations can be applied at the mechanism-level (\pthree).
In the general case, this may be done through sharing less hardware
resources~\cite{Saltzer1975}.  For transient execution side channels, solutions
partly consist in microcode updates to existing hardware~\cite{Canella2019},
fixing existing hardware designs~\cite{Canella2019}, or designing entirely new
hardware mechanisms resilient to side channels~\cite{Kogler2023, Narayan2023}.
Some mitigations to transient execution side channels, e.g., Spectre attacks,
cannot be achieved in hardware.  Approaches employ ad-hoc compiler-based
mitigations~\cite{Schlueter2022}, pointer masking~\cite{Narayan2021, Jin2024},
or combinations of software and hardware mechanisms~\cite{Narayan2021}.  All
come at a sizable performance cost~\cite{Behrens2022} which should be factored
into broader mechanism performance considerations (Insight
\hyperlink{performanceinsight}{13}).

Still, not all mitigations to software-exploitable side channels can be done at
the level of the mechanism.  Other mitigations include designing applications
themselves to break the correlation between side channels and secrets with
constant-time programming techniques~\cite{Barthe2014, Watt2019, Jancar2022}
made robust to transient execution attacks~\cite{Cauligi2020}, or limiting
attackers' measurement abilities through, e.g., reducing the resolution of
timers or designing APIs that impede profiling-like behavior~\cite{Oren2015}.
These too come at a performance cost and must be implemented at the level of
\pone and \ptwo.

\insight{
The problem of side channels spans the entire \ponetothree stack.
Counter-measures in compartmentalization remain in their infancy: new
mechanisms commonly scope out side channels~\cite{Jero2022, Schrammel2022,
Gu2022b, Yuan2023, Amar2023}, and the problem is widely considered as
orthogonal to \pone and \ptwo~\cite{Huang2022, Duy2023, Khan2023,
Peng2023}.  This calls for more research on combined compartmentalization and
side-channels topics to reach side-channel resilience throughout \pone,
\ptwo, and \pthree.
}

%% file: 03-TABLE-P1.tex
\newcommand{\na}[1][]{\textsc{n/a}}
\newcommand{\none}[1][]{$\varnothing$}
\newcommand{\nfirst}[1]{\phantom{\textsuperscript{1}}#1\textsuperscript{1}}
\newcommand{\rnfirst}[1]{#1\makebox[0pt]{ \textsuperscript{1}}}
\newcommand{\ntwo}[1]{\phantom{\textsuperscript{2}}#1\textsuperscript{2}}
\newcommand{\ntwothree}[1]{\phantom{\textsuperscript{2,3}}#1\textsuperscript{2,3}}
\newcommand{\rntwo}[1]{#1\makebox[0pt]{ \textsuperscript{2}}}
\newcommand{\nthree}[1]{\phantom{\textsuperscript{3}}#1\textsuperscript{3}}
\newcommand{\nthreefour}[1]{\phantom{\textsuperscript{3,4}}#1\textsuperscript{3,4}}
\newcommand{\rnthree}[1]{#1\makebox[0pt]{ \textsuperscript{3}}}
\newcommand{\nfour}[1]{\phantom{\textsuperscript{4}}#1\textsuperscript{4}}
\newcommand{\rnfour}[1]{#1\makebox[0pt]{ \textsuperscript{4}}}
\newcommand{\nfive}[1]{\phantom{\textsuperscript{5}}#1\textsuperscript{5}}
\newcommand{\rnfive}[1]{#1\makebox[0pt]{ \textsuperscript{5}}}
\newcommand{\nsix}[1]{\phantom{\textsuperscript{6}}#1\textsuperscript{6}}
\newcommand{\rnsix}[1]{#1\makebox[0pt]{ \textsuperscript{6}}}
\newcommand{\ctric}[1][]{\emph{Code}{-centric}}
\newcommand{\dtric}[1][]{\emph{Data}{-centric}}
\newcommand{\any}[1][]{Any}

\begin{table*}[]
\centering
\caption{\emph{Taxonomy of Policy Definition Methods.} PDMs that also propose an abstraction are marked with {\footnotesize \sdown{}}. Manual ({\footnotesize \manual{}}) and fully automated ({\footnotesize \fullauto{}}) policies do not leverage a policy language, thus the column features \ul{\textsc{n}}ot/\ul{\textsc{a}}pplicable.}
\setlength{\tabcolsep}{3pt}
\scriptsize
\begin{tabular}{|l|c|c|c|c|c|c|c|cc|}
\hline
\multirow{2}{*}{\parbox{2.7cm}{\centering \emph{Policy Definition Method}}} & \multirow{2}{*}{\parbox{1.2cm}{\centering \emph{Automation {\tiny \nooverhead{} \lowoverhead{} \mediumoverhead{} \veryhighoverhead{}}\textsuperscript{1}}}} & \multicolumn{2}{c|}{\emph{Policy Language Type}} & \multirow{2}{*}{\parbox{1.4cm}{\centering \emph{Separation Granularity}}} & \multirow{2}{*}{\parbox{1.3cm}{\centering \emph{Analysis Approach}}} & \multirow{2}{*}{\parbox{1.8cm}{\centering \emph{Subject Selection}}} & \multirow{2}{*}{\parbox{1cm}{\centering \emph{Language Specific}}}  & \multicolumn{2}{c|}{\parbox{3.3cm}{\centering \emph{Additional Goals of Automation}}} \\ \cline{3-4} \cline{9-10}
                                                                  &               & \emph{Annotations} & \emph{Placement Rules} & & & & & \multicolumn{1}{c|}{\emph{Performance}} & \multicolumn{1}{c|}{\emph{Interface Safety}} \\ \hline \hline
Manual~\smlcite{Hsu2016}, [\dots]                                 & \manual           & \na     & \na     & Any          & Manual          & \any   & \Circle     & \multicolumn{1}{c|}{\na}     & \multicolumn{1}{c|}{\na}     \\ \hline \hline
Crowbar~\smlcite{Bittau2008}                                      & \lowautomation    & \Circle & \Circle & Function     & Dynamic         & \ctric & \Circle     & \multicolumn{1}{c|}{\Circle} & \multicolumn{1}{c|}{\Circle} \\ \hline
MPDs\s~\smlcite{Parmer2007, Parmer2012}                           & \lowautomation    & \Circle & \Circle & Component    & Hybrid          & \ctric & \CIRCLE     & \multicolumn{1}{c|}{\CIRCLE} & \multicolumn{1}{c|}{\Circle} \\ \hline
CubicleOS\s~\smlcite{Sartakov2021}                                & \lowautomation    & \CIRCLE & \CIRCLE & $\mu$Library & \ntwo{Static}   & \ctric & \CIRCLE     & \multicolumn{1}{c|}{\Circle} & \multicolumn{1}{c|}{\Circle} \\ \hline
Google SAPI\s~\smlcite{GOOGLE_SAPI}                               & \lowautomation    & \CIRCLE & \CIRCLE & Function     & Static          & \ctric & \CIRCLE     & \multicolumn{1}{c|}{\Circle} & \multicolumn{1}{c|}{\Circle} \\ \hline
FlexOS\s~\smlcite{Lefeuvre2022, Jung2021}                         & \lowautomation    & \CIRCLE & \CIRCLE & $\mu$Library & Dynamic         & \ctric & \CIRCLE     & \multicolumn{1}{c|}{\CIRCLE} & \multicolumn{1}{c|}{\Circle} \\ \hline
RLBox\s~\smlcite{Narayan2020}                                     & \lowautomation    & \CIRCLE & \Circle & Function     & Static          & \any   & \CIRCLE     & \multicolumn{1}{c|}{\Circle} & \multicolumn{1}{c|}{\CIRCLE} \\ \hline
SOAAP\s~\smlcite{Gudka2015}                                       & \lowautomation    & \CIRCLE & \Circle & Any          & Hybrid          & \any   & \CIRCLE     & \multicolumn{1}{c|}{\CIRCLE} & \multicolumn{1}{c|}{\CIRCLE} \\ \hline \hline
SeCage\s~\smlcite{Liu2015}                                        & \mediumautomation & \CIRCLE & \Circle & Function     & Hybrid          & \ctric & \CIRCLE     & \multicolumn{1}{c|}{\Circle} & \multicolumn{1}{c|}{\Circle} \\ \hline
PtrSplit\s~\smlcite{Liu2017}                                      & \mediumautomation & \CIRCLE & \Circle & Function     & Static          & \ctric & \CIRCLE     & \multicolumn{1}{c|}{\Circle} & \multicolumn{1}{c|}{\Circle} \\ \hline
PrivTrans\s~\smlcite{Brumley2004}                                 & \mediumautomation & \CIRCLE & \Circle & Function     & Hybrid          & \ctric & \CIRCLE     & \multicolumn{1}{c|}{\Circle} & \multicolumn{1}{c|}{\Circle} \\ \hline
Glamdring\s~\smlcite{Lind2017}                                    & \mediumautomation & \CIRCLE & \Circle & Function     & Static          & \ctric & \CIRCLE     & \multicolumn{1}{c|}{\Circle} & \multicolumn{1}{c|}{\CIRCLE} \\ \hline
Shreds\s~\smlcite{Chen2016}, {\tiny CAPACITY}\s~\smlcite{Duy2023} & \mediumautomation & \CIRCLE & \Circle & Any          & Static          & \ctric & \Circle     & \multicolumn{1}{c|}{\Circle} & \multicolumn{1}{c|}{\CIRCLE} \\ \hline
DataShield\s~\smlcite{Carr2017}                                   & \mediumautomation & \CIRCLE & \Circle & Any          & Static          & \ctric & \Circle     & \multicolumn{1}{c|}{\Circle} & \multicolumn{1}{c|}{\CIRCLE} \\ \hline
Swift\s~\smlcite{Chong2007}                                       & \mediumautomation & \CIRCLE & \Circle & Any          & Static          & \ctric & \CIRCLE     & \multicolumn{1}{c|}{\Circle} & \multicolumn{1}{c|}{\CIRCLE} \\ \hline
Jif\s~\smlcite{Zdancewic2002}                                     & \mediumautomation & \CIRCLE & \CIRCLE & Any          & Static          & \ctric & \CIRCLE     & \multicolumn{1}{c|}{\Circle} & \multicolumn{1}{c|}{\CIRCLE} \\ \hline
PM~\smlcite{Liu2019a}                                             & \mediumautomation & \CIRCLE & \CIRCLE & Function     & Hybrid          & \ctric & \CIRCLE     & \multicolumn{1}{c|}{\CIRCLE} & \multicolumn{1}{c|}{\Circle} \\ \hline
KSplit\s~\smlcite{Huang2022}                                      & \mediumautomation & \CIRCLE & \CIRCLE & Driver       & Static          & \ctric & \CIRCLE     & \multicolumn{1}{c|}{\Circle} & \multicolumn{1}{c|}{\Circle} \\ \hline
Cali\s~\smlcite{Bauer2021}                                        & \mediumautomation & \Circle & \CIRCLE & Library      & Static          & \ctric & \Circle     & \multicolumn{1}{c|}{\Circle} & \multicolumn{1}{c|}{\Circle} \\ \hline
CompartOS\s~\smlcite{Almatary2022}                                & \mediumautomation & \Circle & \CIRCLE & Linkage Unit & Static          & \ctric & \Circle     & \multicolumn{1}{c|}{\Circle} & \multicolumn{1}{c|}{\Circle} \\ \hline
Enclosure\s~\smlcite{Ghosn2021}                                   & \mediumautomation & \Circle & \CIRCLE & Package      & Static          & \ctric & \CIRCLE     & \multicolumn{1}{c|}{\Circle} & \multicolumn{1}{c|}{\Circle} \\ \hline
BreakApp\s~\smlcite{Vasilakis2018}                                & \mediumautomation & \Circle & \CIRCLE & Package      & Static          & \ctric & \CIRCLE     & \multicolumn{1}{c|}{\Circle} & \multicolumn{1}{c|}{\Circle} \\ \hline
CompARTist\s~\smlcite{Huang2017}                                  & \mediumautomation & \Circle & \CIRCLE & Library      & Static          & \ctric & \CIRCLE     & \multicolumn{1}{c|}{\Circle} & \multicolumn{1}{c|}{\Circle} \\ \hline
ACES\s~\smlcite{Clements2018}                                     & \mediumautomation & \Circle & \CIRCLE & Function     & \nthree{Any}    & \ctric & \Circle     & \multicolumn{1}{c|}{\Circle} & \multicolumn{1}{c|}{\Circle} \\ \hline \hline
ProgramCutter~\smlcite{Wu2013}                                    & \fullauto         & \na     & \na     & Function     & Dynamic         & \ctric & \CIRCLE     & \multicolumn{1}{c|}{\Circle} & \multicolumn{1}{c|}{\Circle} \\ \hline
$\mu$SCOPE~\smlcite{Roessler2021}, {\tiny SCALPEL}~\smlcite{Roessler2021a}& \fullauto & \na     & \na     & Any          & Dynamic         & \ctric & \Circle     & \multicolumn{1}{c|}{\CIRCLE} & \multicolumn{1}{c|}{\Circle} \\ \hline
\multicolumn{10}{r}{\raggedleft \textsuperscript{1} \manual{} = \hyperlink{manual}{manual}, \lowautomation{} = \hyperlink{guidedm}{guided manual}, \mediumautomation{} = \hyperlink{policyr}{policy refinement}, \fullauto{} = \hyperlink{fullauto}{full automation}. \textsuperscript{2} Loader-based. \textsuperscript{3} Implemented with static analysis, dynamic analysis possible~\smlcite{Clements2018}.} \\
\end{tabular}
\label{fig:taxonomy-algos}
\vspace{-0.320cm}
\end{table*}

%% file: 03-TABLE-P2.tex
\newcommand{\knl}[1][]{\textsc{k}}
\newcommand{\uk}[1][]{\textsc{u/k}}
\newcommand{\ukx}[1][]{\textsc{u+k}}
\newcommand{\both}[1][]{$\mathcal{S}$+$\mathcal{A}$}

\begin{table*}[]
\newcommand{\mpd}[1][]{Mutable Protection Domains (MPDs)~\smlcite{Parmer2007, Parmer2012}}
\newcommand{\processes}[1][]{POSIX Processes (and earlier instances)~\smlcite{Daley1968}}
\newcommand{\smv}[1][]{Secure Memory Views (SMVs)~\smlcite{Hsu2016}}
\newcommand{\lwc}[1][]{Light-Weight Contexts (LwCs)~\smlcite{Litton2016}}
\newcommand{\microkernels}[1][]{Microkernel Servers~\smlcite{Elphinstone2013}, {\tiny [\dots]}}
\newcommand{\sapi}[1][]{Google Sandboxed API (SAPI)~\smlcite{GOOGLE_SAPI}}
\newcommand{\lvds}[1][]{LVDs / KSplit~\smlcite{Narayanan2020,Huang2022}}
\newcommand{\rlbox}[1][]{RLBox / $\mu$\textsc{switch}~\smlcite{Narayan2020, Peng2023}}
\newcommand{\ptrsplit}[1][]{PtrSplit / Program Mandering~\smlcite{Liu2017, Liu2019a}}

\centering
\caption{\emph{Taxonomy of Compartmentalization Abstractions.} \emph{Targets}: \ul{\textsc{u}}ser, \ul{\textsc{k}}ernel, \ul{\textsc{h}}yper\ul{\textsc{v}}isor. \emph{Semantics}: \ul{$\mathcal{S}$}ynchronous, \ul{$\mathcal{A}$}synchronous, \ul{\textsc{sh}}ared \ul{\textsc{m}}emory, \ul{\textsc{mes}}sage passing. \both: the abstraction exhibits both $\mathcal{S}$ and $\mathcal{A}$ semantics. \emph{Properties}: \ul{\textsc{c}}onfidentiality, \ul{\textsc{i}}ntegrity, \ul{\textsc{a}}vailability, \ul{\textsc{r}}ecompartmentalization. Mechanism-independent abstractions are labeled with \none.}
\setlength{\tabcolsep}{3pt}
\scriptsize
\begin{tabular}{|cc|c|l|c|c|c|c|l|l|l|l|c|c|}
\hline
\multicolumn{2}{|c|}{\multirow{2}{*}{\emph{Class}}} & \emph{Target} & \multicolumn{1}{c|}{\multirow{2}{*}{\emph{Abstraction}}} & \multicolumn{1}{c|}{\multirow{2}{*}{\emph{Subject Selection}}} & \multicolumn{2}{c|}{\multirow{1}{*}{\emph{Semantics}}} & \multirow{2}{*}{\parbox{1.5cm}{\centering \emph{Abstraction Granularity}}} & \multicolumn{4}{c|}{\emph{Properties}} & \multirow{2}{*}{\parbox{0.9cm}{\centering \emph{Interface Safety}}} & \multirow{2}{*}{\parbox{1.6cm}{\centering \emph{Design Bound to Mechanism}}} \\ \cline{6-7} \cline{9-12}
& & \textsc{u/k/hv} & & & \tinytt{CALL} & \tinytt{ASSIGN} & & \multicolumn{1}{c|}{\textsc{c}} & \multicolumn{1}{c|}{\textsc{i}} & \multicolumn{1}{c|}{\textsc{a}} & \multicolumn{1}{c|}{\textsc{r}} & & \\ \hline \hline
\multicolumn{2}{|c|}{\multirow{22}{*}{\rotatebox{90}{\emph{Mutual Distrust}}}}
  & \multirow{12}{*}{\textsc{u}} & Virtines~\smlcite{Wanninger2022}           & \ctric & $\mathcal{S}$ & \textsc{mes} & Function          & \CIRCLE & \CIRCLE & \Circle     & \Circle & \Circle          & Virtual Machine (EPT) \\ \cline{4-14}
                             & & & ACES~\smlcite{Clements2018}                & \ctric & $\mathcal{S}$ & \textsc{shm} & Function          & \CIRCLE & \CIRCLE & \Circle     & \Circle & \Circle          & \nfive{\none}         \\ \cline{4-14}
                             & & & SeCage~\smlcite{Liu2015}                   & \ctric & $\mathcal{S}$ & \textsc{shm} & Function          & \CIRCLE & \CIRCLE & \Circle     & \Circle & \Circle          & \nfive{\none}         \\ \cline{4-14}
			     & & & HODOR~\smlcite{Hedayati2019}               & \ctric & $\mathcal{S}$ & \textsc{shm} & Library           & \CIRCLE & \CIRCLE & \Circle     & \Circle & \Circle          & \none                 \\ \cline{4-14}
			     & & & \textsc{capacity}~\smlcite{Duy2023}        & \ctric & $\mathcal{S}$ & \textsc{shm} & \any              & \CIRCLE & \CIRCLE & \Circle     & \Circle & \nfour{\Circle}  & ARM PAC + MTE         \\ \cline{4-14}
			     & & & Jif~\smlcite{Zdancewic2002}                & \ctric & \both         & \textsc{mes} & \any              & \CIRCLE & \CIRCLE & \Circle     & \Circle & \CIRCLE          & \none                 \\ \cline{4-14}
			     & & & Arbiter~\smlcite{Wang2015}                 & \dtric & $\mathcal{S}$ & \textsc{shm} & \nfirst{Function} & \CIRCLE & \CIRCLE & \Circle     & \Circle & \Circle          & \nfive{\none}         \\ \cline{4-14}
			     & & & \smv                                       & \dtric & $\mathcal{S}$ & \textsc{shm} & \nfirst{Function} & \CIRCLE & \CIRCLE & \Circle     & \Circle & \Circle          & \nfive{\none}         \\ \cline{4-14}
			     & & & Salus~\smlcite{Strackx2015}                & \dtric & $\mathcal{S}$ & \textsc{shm} & \nfirst{Function} & \CIRCLE & \CIRCLE & \Circle     & \Circle & \Circle          & \nfive{\none}         \\ \cline{4-14}
			     & & & \lwc                                       & Hybrid & $\mathcal{S}$ & \textsc{shm} & \nfirst{Function} & \CIRCLE & \CIRCLE & \Circle     & \Circle & \Circle          & \ntwo{Page Table}     \\ \cline{4-14}
			     & & & \processes                                 & Hybrid & \any          & \any         & \any              & \CIRCLE & \CIRCLE & \Circle     & \Circle & \Circle          & Page Table            \\ \cline{4-14}
                             & & & SOAAP~\smlcite{Gudka2015}                  & Hybrid & $\mathcal{S}$ & \textsc{shm} & \any              & \CIRCLE & \CIRCLE & \Circle     & \Circle & \nfour{\Circle}  & \none                 \\ \cline{4-14}
                             & & & libMPK~\smlcite{Park2019}                  & Hybrid & $\mathcal{S}$ & \textsc{shm} & \any              & \CIRCLE & \CIRCLE & \Circle     & \Circle & \Circle          & Protection Keys       \\ \cline{3-14}
     & & \multirow{7}{*}{\ukx{}} & CheriOS~\smlcite{Esswood2021}              & \ctric & \any          & \any         & \uk{}-component   & \CIRCLE & \CIRCLE & \Circle     & \Circle & \Circle          & CHERI                 \\ \cline{4-14}
                             & & & \microkernels                              & \ctric & \any          & \textsc{mes} & \uk{}-component   & \CIRCLE & \CIRCLE & \Circle     & \Circle & \Circle          & \nfive{\none}         \\ \cline{4-14}
			     & & & \mpd{}                                     & \ctric & $\mathcal{S}$ & \textsc{shm} & \uk{}-component   & \CIRCLE & \CIRCLE & \Circle     & \CIRCLE & \Circle          & \nfive{\none}         \\ \cline{4-14}
                             & & & RedLeaf~\smlcite{Narayanan2020a}           & \ctric & $\mathcal{S}$ & \textsc{shm} & \uk{}-component   & \CIRCLE & \CIRCLE & \CIRCLE     & \Circle & \Circle          & Safe Languages        \\ \cline{4-14}
			     & & & CubicleOS~\smlcite{Sartakov2021}           & \ctric & $\mathcal{S}$ & \textsc{shm} & $\mu$Library      & \CIRCLE & \CIRCLE & \Circle     & \Circle & \Circle          & Protection Keys       \\ \cline{4-14}
                             & & & FlexOS~\smlcite{Lefeuvre2022}              & \ctric & $\mathcal{S}$ & \textsc{shm} & $\mu$Library      & \CIRCLE & \CIRCLE & \Circle     & \Circle & \Circle          & \none                 \\ \cline{4-14}
                             & & & xMP~\smlcite{Proskurin2020}                & \ctric & $\mathcal{S}$ & \textsc{shm} & \any              & \CIRCLE & \CIRCLE & \Circle     & \Circle & \Circle          & \none                 \\ \cline{4-14}
			     & & & Monza~\smlcite{MONZA}                      & Hybrid & $\mathcal{A}$ & \textsc{shm} & \nfirst{Function} & \Circle & \CIRCLE & \Circle     & \Circle & \Circle          & \nfive{\none}         \\ \cline{3-14}
     & & \multirow{3}{*}{\knl{}} & VirtuOS~\smlcite{Nikolaev2013}             & \ctric & \both         & \textsc{shm} & \knl{}-component  & \CIRCLE & \CIRCLE & \CIRCLE     & \Circle & \Circle          & Virtual Machine (EPT) \\ \cline{4-14}
                             & & & HAKC~\smlcite{McKee2022}                   & \ctric & $\mathcal{S}$ & \textsc{shm} & Function          & \CIRCLE & \CIRCLE & \Circle     & \Circle & \Circle          & ARM PAC + MTE         \\ \cline{4-14}
                             & & & LibrettOS~\smlcite{Nikolaev2020}           & \ctric & $\mathcal{S}$ & \textsc{shm} & \knl{}-component  & \CIRCLE & \CIRCLE & \CIRCLE     & \CIRCLE & \Circle          & \nfive{\none}         \\ \hline \hline
\multicolumn{2}{|c|}{\multirow{10}{*}{\rotatebox{90}{\emph{Sandbox}}}}
   & \multirow{6}{*}{\textsc{u}} & Cali~\smlcite{Bauer2021}                   & \ctric & $\mathcal{S}$ & \textsc{shm} & Library           & \CIRCLE & \CIRCLE & \Circle     & \Circle & \Circle          & \nfive{\none}         \\ \cline{4-14}
                             & & & CompARTist~\smlcite{Huang2017}             & \ctric & $\mathcal{S}$ & \textsc{mes} & Library           & \CIRCLE & \CIRCLE & \Circle     & \Circle & \Circle          & \nfive{\none}         \\ \cline{4-14}
			     & & & Enclosure~\smlcite{Ghosn2021}              & \ctric & $\mathcal{S}$ & \textsc{shm} & Package           & \CIRCLE & \CIRCLE & \Circle     & \Circle & \Circle          & \none                 \\ \cline{4-14}
			     & & & \sapi                                      & \ctric & $\mathcal{S}$ & \textsc{mes} & Function          & \CIRCLE & \CIRCLE & \CIRCLE     & \Circle & \Circle          & \nfive{\none}         \\ \cline{4-14}
	                     & & & \rlbox                                     & Hybrid & $\mathcal{S}$ & \textsc{shm} & Function          & \CIRCLE & \CIRCLE & \Circle     & \Circle & \CIRCLE          & \none                 \\ \cline{4-14}
			     & & & Wedge~\smlcite{Bittau2008}                 & Hybrid & $\mathcal{S}$ & \textsc{shm} & \nfirst{Function} & \CIRCLE & \CIRCLE & \Circle     & \Circle & \Circle          & \nfive{\none}         \\ \cline{3-14}
                      & & \ukx{} & CompartOS~\smlcite{Almatary2022}           & \ctric & $\mathcal{S}$ & \textsc{shm} & Linkage Unit      & \CIRCLE & \CIRCLE & \CIRCLE     & \Circle & \Circle          & CHERI                 \\ \cline{3-14}
     & & \multirow{2}{*}{\knl{}} & \lvds                                      & \ctric & $\mathcal{S}$ & \textsc{mes} & \knl{}-component  & \CIRCLE & \CIRCLE & \Circle     & \Circle & \Circle          & \nfive{\none}         \\ \cline{4-14}
                             & & & XFI/LXFI~\smlcite{Erlingsson2006, Mao2011} & Hybrid & $\mathcal{S}$ & \textsc{shm} & \knl{}-component  & \CIRCLE & \CIRCLE & \Circle     & \Circle & \CIRCLE          & SFI                   \\ \cline{3-14}
		 & & \textsc{hv} & Nexen~\smlcite{Shi2017}                    & \dtric & $\mathcal{S}$ & \textsc{shm} & Per-VM domain     & \CIRCLE & \CIRCLE & \Circle     & \Circle & \Circle          & \nthree{Page Table}   \\ \hline \hline
\multicolumn{1}{|c}{\multirow{9}{*}{\rotatebox{90}{\parbox{1.2cm}{\centering \emph{Safebox}}}}}
& & \multirow{1}{*}{\textsc{u}} & Shreds~\smlcite{Chen2016}                   & \ctric & $\mathcal{S}$ & \textsc{shm} & \any              & \CIRCLE & \CIRCLE & \Circle     & \Circle & \nfour{\Circle}  & \nfive{\none}         \\ \cline{2-14}
& \multicolumn{1}{|c|}{\multirow{8}{*}{\rotatebox{90}{\parbox{1.2cm}{\centering \emph{Dual World}}}}} & \multirow{7}{*}{\textsc{u}}
                                 & Privman~\smlcite{Kilpatrick2003}           & \ctric & $\mathcal{S}$ & \textsc{mes} & Function          & \CIRCLE & \CIRCLE & \Circle     & \Circle & \Circle          & \ntwo{Page Table}     \\ \cline{4-14}
      & \multicolumn{1}{|c|}{} & & Privtrans~\smlcite{Brumley2004}            & \ctric & $\mathcal{S}$ & \textsc{mes} & Function          & \CIRCLE & \CIRCLE & \Circle     & \Circle & \Circle          & \ntwo{Page Table}     \\ \cline{4-14}
      & \multicolumn{1}{|c|}{} & & Swift~\smlcite{Chong2007}                  & \ctric & \both         & \textsc{mes} & \any              & \CIRCLE & \CIRCLE & \Circle     & \Circle & \CIRCLE          & \none                 \\ \cline{4-14}
      & \multicolumn{1}{|c|}{} & & Glamdring~\smlcite{Lind2017}               & \ctric & $\mathcal{S}$ & \textsc{mes} & Function          & \CIRCLE & \CIRCLE & \Circle     & \Circle & \CIRCLE          & \nfive{\none}         \\ \cline{4-14}
      & \multicolumn{1}{|c|}{} & & \ptrsplit                                  & \ctric & $\mathcal{S}$ & \textsc{mes} & Function          & \CIRCLE & \CIRCLE & \Circle     & \Circle & \nfour{\Circle}  & \nfive{\none}         \\ \cline{4-14}
      & \multicolumn{1}{|c|}{} & & DataShield~\smlcite{Carr2017}              & Hybrid & $\mathcal{S}$ & \textsc{shm} & \any              & \CIRCLE & \CIRCLE & \Circle     & \Circle & \nfour{\Circle}  & Bounds Checking       \\ \cline{4-14}
      & \multicolumn{1}{|c|}{} & & ERIM~\smlcite{VahldiekOberwagner2019}      & Hybrid & $\mathcal{S}$ & \textsc{shm} & \any              & \CIRCLE & \CIRCLE & \Circle     & \Circle & \Circle          & Protection Keys       \\ \cline{3-14}
& \multicolumn{1}{|c|}{} & \knl{}& Nested Kernel~\smlcite{Dautenhahn2015}     & \ctric & $\mathcal{S}$ & \textsc{shm} & Function          & \CIRCLE & \CIRCLE & \Circle     & \Circle & \Circle          & Page Table            \\ \hline
\multicolumn{14}{r}{\hfill \textsuperscript{1} Inherited from thread-like semantics, \textsuperscript{2} from process-like semantics, \textsuperscript{3} from the Nested Kernel, \textsuperscript{4} The PDM does (to a certain extent).} \\
\multicolumn{14}{r}{\raggedleft \textsuperscript{5} The abstraction could plug onto any intra-AS mechanism, though the paper or documentation claims reliance on a particular one.} \\
\end{tabular}
\label{fig:abstraction-taxonomy}
\vspace{-0.3cm}
\end{table*}

% - TODO RedLeaf should probably have the interface safety marked!
% - TODO Enclosure might need to be in mutual distrust
% - TODO CompartOS should probably be in the S + S category
% - TODO DataShield and Glamdring should probably have N/A in ASSIGN semantics since they are isolation abstractions

%% file: 03-TABLE-P3.tex
\begin{table*}[]
% these are some entries of the table that we moved out to make the tex file more readable
\newcommand{\sfi}[1][]{Software Fault Isolation~\smlcite{Wahbe1993, Castro2009, Yee2009, Zeng2011, Morrisett2012, Haas2017, Johnson2021, Johnson2023}}
\newcommand{\ab}[1][]{Access Bits~\smlcite{INTEL_PT_PROTECTION}}
\newcommand{\cvms}[1][]{Confidential VMs~\smlcite{INTEL_TDX, AMD_SEV, ARM_CCA}}
\newcommand{\ept}[1][]{EPT / \tinytt{vmfunc}~\smlcite{INTEL_EPT}}
\newcommand{\eptov}[1][]{\overhead{360} (PT switch + \textsuperscript{5})}
\newcommand{\hwcapov}[1][]{\overhead{180} -- \overhead{360} (impl. dep., incl. \textsuperscript{5})}
\newcommand{\sbcov}[1][]{\overhead{180} -- \overhead{360} (impl. dep., incl. \textsuperscript{5})}
\newcommand{\enclaves}[1][]{Enclaves~\smlcite{INTEL_SGX, Costan2016a}}
\newcommand{\pkeys}[1][]{Protection Keys~\smlcite{IBM_ZOS_PKEYS, INTEL_MPK_SDM, ARM_MPU, Schrammel2020, Yuan2023}}
\newcommand{\pkeygran}[1][]{8-1024~\smlcite{ARM_MPU, Schrammel2020}\textsuperscript{4}}
\newcommand{\pkeyov}[1][]{\overhead{90} (special register flip + \textsuperscript{5})}
\newcommand{\sflang}[1][]{Safe Languages~\smlcite{RUST_LANG}}
\newcommand{\vflang}[1][]{Software Verification~\smlcite{Leino2010, Kolosick2022}}
\newcommand{\bchw}[1][]{Bounds-Checking Hardware~\smlcite{Devietti2008, INTEL_MPX, Sasaki2019, Kim2020, Koning2017}}
\newcommand{\mte}[1][]{(Other) Tagged Architectures~\smlcite{ARM_MTE, SPARC_ADI, Song2016, Weiser2019, Hu2021a, Roessler2018, Jero2022, Dhawan2015}}
\newcommand{\mteov}[1][]{\overhead{180} (tagging hardware + \textsuperscript{5})}
\newcommand{\mtegran}[1][]{16\textsuperscript{4} - $\infty$~\smlcite{Jero2022}}
\newcommand{\hwcap}[1][]{Hardware Capabilities~\smlcite{Carter1994, Vilanova2014, Watson2015, Nam2019, Amar2023}}
\newcommand{\swcap}[1][]{Software Capabilities~\smlcite{Heiser1999, Chase1994}}
\newcommand{\segm}[1][]{Segmentation-like Hardware~\smlcite{Frassetto2018, Mogosanu2018}}
\newcommand{\mondriaan}[1][]{Mondrian Memory Protection (MMP)~\smlcite{Witchel2002}}
\newcommand{\worldsep}[1][]{World Separation~\smlcite{ARM_TRUSTZONE, AMD_PSP}}
\newcommand{\superbit}[1][]{Supervisor Bit~\smlcite{INTEL_PT_PROTECTION, Lee2018, Lee2023}}
\newcommand{\fcircle}[1][]{\rntwo{\CIRCLE}}

\centering
\caption{\emph{Taxonomy of Compartmentalization Mechanisms.} Page-Table = PT; Permissions: \ul{\textsc{r}}ead, \ul{\textsc{w}}rite, \ul{\textsc{e}}xecute, \ul{\textsc{a}}ddress (\emph{create pointers to}), \CIRCLE{} = supported, \LEFTcircle{} = supported by some, \Circle{} = unsupported; Overhead: \emph{free}{\scriptsize{}\hspace{0.05cm}=\hspace{0.05cm}\nooverhead{}\hspace{0.05cm}$<$\hspace{0.05cm}\lowoverhead{}\hspace{0.05cm}$<$\hspace{0.05cm}\mediumoverhead{}\hspace{0.05cm}$<$\hspace{0.05cm}\highoverhead{}\hspace{0.05cm}$<$\hspace{0.05cm}\veryhighoverhead{}\hspace{0.05cm}=\hspace{0.05cm}}\emph{very costly}.}

% tricks to reduce the size of the table
\setlength{\tabcolsep}{3pt}
\scriptsize

\begin{tabular}{|c|cl|c|l|l|l|l|l|l|l|c|l|}
\hline
% headers
\multicolumn{1}{|c|}{} & \multicolumn{2}{c|}{\multirow{2}{*}{\textit{Mechanism Class}}} & \multicolumn{1}{c|}{\multirow{2}{*}{\parbox{0.8cm}{\centering \textit{Condi-\\tioned}}}} & \multicolumn{1}{c|}{\multirow{2}{*}{\parbox{0.7cm}{\centering \textit{Trust\\Model}}}}
   & \multicolumn{1}{c|}{\multirow{2}{*}{\textit{TCB}}} & \multicolumn{4}{c|}{\textit{Permissions}} & \multicolumn{1}{c|}{\multirow{2}{*}{\textit{Granularity}}} & \multicolumn{1}{c|}{\multirow{2}{*}{\textit{\textnumero{} of Domains}}}
   & \multicolumn{1}{c|}{\multirow{1}{*}{\textit{Domain Switch Cost}}} \\ \cline{7-10}
& & & & & & \multicolumn{1}{c|}{\textsc{r}} & \multicolumn{1}{c|}{\textsc{w}} & \multicolumn{1}{c|}{\textsc{x}} & \multicolumn{1}{c|}{\textsc{a}} & & & \multicolumn{1}{c|}{\scriptsize{\textit{(Versus Non-Separated)}}} \\ \hline \hline
% table content
\multicolumn{1}{|c|}{\multirow{13}{*}{\rotatebox{90}{Hardware}}}
  & \multicolumn{2}{l|}{Physical Separation~\smlcite{Rushby1981}}        & \Circle                & Mutual & Full & \CIRCLE    & \CIRCLE    & \CIRCLE     & \Circle & Physical Mem.                      & \textnumero{} of machines  & \overhead{180} -- \overhead{360} (link latency)            \\ \cline{2-13}
  & \multicolumn{1}{c|}{\multirow{4}{*}{PT}} & \ab{}, \ept{}             & \nfirst{\LEFTcircle{}} & Mutual & Full & \fcircle{} & \fcircle{} & \fcircle{}  & \Circle & Page                               & $\infty$                   & \eptov{}                                                   \\ \cline{3-13}
    & \multicolumn{1}{c|}{} & \superbit{}                                & \Circle                & Single & Full & \fcircle{} & \fcircle{} & \fcircle{}  & \Circle & Page                               & 2 (kernel/user)            & \overhead{270} (interrupt + \textsuperscript{5})           \\ \cline{3-13}
    & \multicolumn{1}{c|}{} & \mondriaan{}                               & \Circle                & Mutual & Full & \fcircle{} & \fcircle{} & \fcircle{}  & \Circle & Word                               & $\infty$                   & \overhead{180} (MMP hardware + \textsuperscript{5})        \\ \cline{3-13}
    & \multicolumn{1}{c|}{} & \pkeys{}                                   & \CIRCLE                & Mutual & Full & \CIRCLE    & \CIRCLE    & \LEFTcircle & \Circle & Page                               & \pkeygran{}                & \pkeyov{}                                                  \\ \cline{2-13}
  & \multicolumn{2}{l|}{\segm{}}                                         & \Circle                & Single & Full & \CIRCLE    & \CIRCLE    & \Circle     & \Circle & Byte - Page~\smlcite{Mogosanu2018} & 2 (safe/unsafe)            & \overhead{90} (\textsuperscript{5})                        \\ \cline{2-13}
  & \multicolumn{1}{c|}{\multirow{3}{*}{TEE}} & \enclaves{}              & \Circle                & Mutual & TEE  & \CIRCLE    & \CIRCLE    & \CIRCLE     & \Circle & Page                               & $\infty$                   & \overhead{360} (enclave call, incl. \textsuperscript{5})   \\ \cline{3-13}
    & \multicolumn{1}{c|}{} & \cvms{}                                    & \Circle                & Mutual & TEE  & \CIRCLE    & \CIRCLE    & \CIRCLE     & \Circle & Page                               & $\infty$                   & \overhead{360} ($>$ EPT switch)                            \\ \cline{3-13}
    & \multicolumn{1}{c|}{} & \worldsep{}                                & \Circle                & Single & TEE  & \CIRCLE    & \CIRCLE    & \CIRCLE     & \Circle & Page                               & 2 (trusted/rest)           & \overhead{360} (world switch, incl. \textsuperscript{5})   \\ \cline{2-13}
  & \multicolumn{2}{l|}{\hwcap{}}                                        & \Circle                & Mutual & Full & \CIRCLE    & \CIRCLE    & \CIRCLE     & \CIRCLE & Byte                               & $\infty$                   & \overhead{180} (special instr. + \textsuperscript{5})      \\ \cline{2-13}
  & \multicolumn{2}{l|}{\bchw{}}                                         & \CIRCLE                & Mutual & Full & \fcircle{} & \fcircle{} & \Circle     & \Circle & Byte                               & $\infty$                   & \overhead{180} (bounds hardware + \textsuperscript{5})     \\ \cline{2-13}
  & \multicolumn{2}{l|}{\mte{}}                                          & \CIRCLE                & Mutual & Full & \fcircle{} & \fcircle{} & \Circle     & \Circle & Byte - Words\textsuperscript{3}    & \mtegran{}                 & \mteov{}                                                   \\ \hline \hline
\multicolumn{1}{|c|}{\multirow{5}{*}{\rotatebox{90}{Software}}}
  & \multicolumn{2}{l|}{\swcap{}}                                        & \Circle                & Mutual & Full & \CIRCLE    & \CIRCLE    & \CIRCLE     & \CIRCLE & Byte                               & $\infty$                   & \hwcapov{}                                                 \\ \cline{2-13}
  & \multicolumn{2}{l|}{Bounds-Checking Software~\smlcite{Song2019}}     & \CIRCLE                & Mutual & Full & \CIRCLE    & \CIRCLE    & \LEFTcircle & \Circle & Byte                               & $\infty$                   & \sbcov{}                                                   \\ \cline{2-13}
  & \multicolumn{2}{l|}{\sflang{} / \vflang{}}                           & \Circle                & Single & Full & \CIRCLE    & \CIRCLE    & \CIRCLE     & \CIRCLE & Byte                               & 2 (safe/unsafe)            & \overhead{0} (function call)                               \\ \cline{2-13}
  & \multicolumn{2}{l|}{\sfi{}}                                          & \Circle                & Single & Full & \CIRCLE    & \CIRCLE    & \CIRCLE     & \CIRCLE & Byte                               & $\infty$                   & \overhead{90} (\textsuperscript{5})                        \\ \cline{2-13}
  & \multicolumn{2}{l|}{Memory Encryption / AES-NI~\smlcite{Koning2017}} & \Circle                & Mutual & Full & \CIRCLE    & \CIRCLE    & \Circle     & \Circle & 128 bits                           & $\infty$                   & \highoverhead{} (copy key + encrypt + \textsuperscript{5}) \\ \hline
% footers
\multicolumn{13}{l}{\hfill \textsuperscript{1} In Ring 0. \textsuperscript{2} Not all combinations of \textsc{r}/\textsc{w}/\textsc{x} supported. \textsuperscript{3} Covers many granularities~\smlcite{Jero2022}. \textsuperscript{4} Some works~\smlcite{Park2019, Gu2022b, McKee2022} increase it. \textsuperscript{5} Register saving/scrubbing, stack switch.} \\
\end{tabular}
\label{fig:taxonomy-mechanisms}
\vspace{-0.3cm}
\end{table*}

% Note: why is bounds-checking conditioned?
% With only bounds-checking hardware or software, we cannot enforce compartment
% entry points. An additional defense such as CFI is needed.

%% file: 04-deployed.tex
\section{Deployed Compartmentalized Software}\label{sec:deployed}
\label{webserver-sep}

% NOTE from Hugo: Postgres is *not* sandboxed or compartmentalized. It has a
% user system, yes, a lot of process separation, yes, but not
% compartmentalization or sandboxing. All processes are running with the same
% user, without seccomp / syscall filtering, and the interface among them is
% not hardened.

% NOTE from Hugo: HAProxy is *not* compartmentalized. There is no privilege
% difference between the master and the workers, the feature is more there
% for threading reasons. HAProxy rather relies on privilege drop at the scale
% of the entire application.

% NOTE from Hugo: excellent document on Dovecot:
% https://doc.dovecot.org/developer_manual/design/processes/#dovecot-processes

\newcommand*{\secu}{Sec. Pro.\xspace}
\newcommand*{\aca}{Academic\xspace}
\newcommand*{\other}{Other\xspace}

\begin{table*}[]
\caption{\emph{Characteristics of Mainstream Compartmentalized Software.} Abbreviations are the same as in \Cref{fig:taxonomy-algos,fig:abstraction-taxonomy,fig:taxonomy-mechanisms}.}
\setlength{\tabcolsep}{2.5pt}
\scriptsize
\centering
\begin{tabular}{|c|c|c|c|c|c|c|c|c|c|c|}
\hline
\multirow{1}{*}{\textit{Software Class}} & \multirow{1}{*}{\textit{Author}} & \multirow{1}{*}{\textit{PDM}} & \multicolumn{6}{c|}{\textit{Abstraction (\S\ref{p2-taxonomy})}} & \multirow{1}{*}{\textit{Mechanism}} \\ \cline{4-9}
\multirow{1}{*}{\textit{(Full list in Appendix \hyperlink{sourcesearchappendix}{A2})}} & \nfirst{\emph{Profile}} & (\S\ref{automation-tradeoffs}) & \multicolumn{1}{c|}{\emph{Name}} & \multicolumn{1}{c|}{\emph{\hyperlink{tmodel}{Trust Model}}} & \multicolumn{1}{c|}{\emph{\hyperlink{smodel}{Subjects}}} & \multicolumn{1}{c|}{\emph{\hyperlink{semant}{Semantics}}} & \multicolumn{1}{c|}{\emph{\hyperlink{props}{Properties}}} & \multicolumn{1}{c|}{\emph{\hyperlink{granty}{Granularity}}} &  (\S\ref{subsec:taxonomy-p3}) \\ \hline \hline
\parbox{4.2cm}{\centering Google V8~\smlcite{V8_SANDBOX}} & \secu & \multirow{8}{*}{\nthree{\manual{}}} & Custom & \multirow{2}{*}{Sandbox} & \ctric & $\mathcal{S}$, \textsc{shm} & \textsc{ci} & \multirow{12}{*}{\parbox{1.5cm}{\centering Coarse (High-level components)}} & \emph{Fixed}: SFI \\ \cline{1-2} \cline{4-4} \cline{6-8} \cline{10-10}
\parbox{4.2cm}{\centering Browser Site Isolation~\smlcite{Chen2007, Wang2009, Reis2019}} & \secu & & Site Isolation & & \dtric & $\mathcal{A}$, \emph{both} & \textsc{cia} & & \multirow{10}{*}{\parbox{1.2cm}{\centering \emph{Fixed}: PT\\\hfill \break(Due to a dependency to \tinytt{fork()} semantics)}} \\ \cline{1-2} \cline{4-8}
\parbox{4.2cm}{\centering OpenBSD Privilege-Separated\\Userland ($>$40 apps\textsuperscript{2} incl. OpenSSH)~\smlcite{OPENBSD_SEP}} & \secu & & \multirow{9}{*}{\parbox{1.35cm}{\centering Custom\\(Process-\\Based)}} & (\emph{mainly}) Sandbox & \parbox{1.35cm}{\centering (\emph{mainly}) \ctric} & $\mathcal{S}$, \textsc{mes} & \textsc{ci(a)} &  & \\ \cline{1-2} \cline{5-8}
\parbox{4.2cm}{\centering IRSSI~\smlcite{IRSSI}} & \aca & & & Safebox & \ctric & $\mathcal{S}$, \textsc{mes} & \textsc{ci} & & \\ \cline{1-2} \cline{5-8}
\parbox{4.2cm}{\centering Debian \tinytt{man}~\smlcite{MANDB}} & \other & & & \multirow{7}{*}{Sandbox} & \ctric & $\mathcal{S}$, \textsc{mes} & \textsc{ci} & & \\ \cline{1-2} \cline{6-8}
\parbox{4.2cm}{\centering Wireshark~\smlcite{WIRESHARK_PRIVSEP}} & \secu & & & & \ctric & $\mathcal{A}$, \textsc{mes} & \textsc{ci} & & \\ \cline{1-2} \cline{6-8}
\parbox{4.2cm}{\centering DHCPCD~\smlcite{DHCPCD}} & \other & & & & \dtric & $\mathcal{A}$, \textsc{mes} & \textsc{cia} & & \\ \cline{1-2} \cline{3-3} \cline{6-8}
\parbox{4.2cm}{\centering VSFTPD~\smlcite{VSFTPD}} & \secu & \multirow{5}{*}{\manual{}} & & & \dtric & $\mathcal{S}$, \textsc{mes} & \textsc{ci} & & \\ \cline{1-2} \cline{6-8}
\parbox{4.2cm}{\centering qmail~\smlcite{Bernstein2007,Hafiz2004}, Postfix~\smlcite{Hafiz2008}, djbdns~\smlcite{Wright2008}} & \secu & & & & \ctric & $\mathcal{A}$, \textsc{mes} & \textsc{cia} & & \\ \cline{1-2} \cline{6-8}
\parbox{4.2cm}{\centering Dovecot~\smlcite{DOVECOT_PRIVSEP}} & \other & & & & \ctric & $\mathcal{A}$, \textsc{mes} & \textsc{cia} & & \\ \cline{1-2} \cline{6-8}
\parbox{4.2cm}{\centering Web Servers~\smlcite{NGINX_ARCH, APACHE_ARCH, LIGHTTPD_ARCH}} & \other & & & & \dtric & $\mathcal{A}$, \emph{both} & \textsc{cia} & & \\ \cline{1-2} \cline{4-8}  \cline{10-10}
Microkernels~\smlcite{Young1987, Liedtke1995, Herder2006, Klein2009, Elphinstone2013} {\tiny [\dots]} & \aca & & Microkernel & Mutual Distrust & \ctric & \emph{both}, \textsc{mes} & \textsc{cia} & & \nfive{\emph{Fixed}: PT}\\ \hline
\parbox{4.2cm}{\centering Firefox Library Sandboxing~\smlcite{MOZ_RLBOX}} & \aca & \nthreefour{\lowautomation{}} & RLBox & Sandbox & \ctric & $\mathcal{S}$, \textsc{shm} & \textsc{cia} & \parbox{1.7cm}{\centering Finer (Libraries)} & \emph{Flexible} \\ \hline
\multicolumn{10}{r}{\hfill \textsuperscript{1}\secu = Security Professional, \textsuperscript{2}Including OpenSSH/NTPd/SMTPd, and others -- see Appendix \hyperlink{sourcesearchappendix}{A2}, \textsuperscript{3}Separation was retrofitted, \textsuperscript{4}RLBox, \textsuperscript{5}Alternatives in research~\smlcite{Gu2020}.} \\
\end{tabular}
\label{tab:char-mainstream-soft}
\vspace{-0.35cm}
\end{table*}

\emph{How does the vast state of the art systematized in
\S\ref{sec:stateoftheart} translate in practice?} To answer this, we discuss a
corpus of \totalfound mainstream compartmentalized programs
(\Cref{tab:char-mainstream-soft}).  We constitute the corpus via a systematic
search in the Debian archive~\cite{DEBIAN_REPO} (\totalsystemfound apps),
completed with previous works (\S\ref{sec:stateoftheart}, \rworksfound
programs), and our own expertise (\expertfound programs).  For the former, we
manually triage all Debian packages with $>$1K installations~\cite{POPCON}
(\popconpackages{} apps), whose source-code feature privilege-separation
keywords.  For clarity we group programs in \classesnum classes, characterized
using our taxonomy (\S\ref{sec:stateoftheart}).  Interested readers can find
the full list of keywords and programs in Appendix
\hyperlink{sourcesearchappendix}{A2}.  We present our insights next.

\customparheader{Software compartmentalization is (still) not the norm.}
As \Cref{tab:char-mainstream-soft} shows, compartmentalized designs started
gaining mainstream awareness in the mid-2000s with programs such as
qmail~\cite{Bernstein2007,Hafiz2004}, OpenSSH~\cite{Provos2003}, or
Postfix~\cite{Hafiz2008}.  Compartmentalization progressed
since then, driven by the challenges of the web, as well as by the OpenBSD and
academic communities.  Still, today compartmentalized designs remain a minority
($<$\totalsystemfound{} out of \popconpackages{} apps), tied to security-aware
vendors (8 / \classesnum{} classes in \Cref{tab:char-mainstream-soft} are
authored by academics or security professionals).  \emph{Non-expert developers, even
of popular software, still do not commonly compartmentalize}.

\customparheader{With skills and time, retrofitting is realistic.}
Partly due to the OpenBSD community, cases where separation was retrofitted
outnumber those architected with separation in mind
(\Cref{tab:char-mainstream-soft}).  This shows that retrofitting is realistic
even in service-critical, long-established codebases such as OpenSSH, V8, or
Firefox. In cases, the deployment of compartmentalization schemes caused
functional regressions, \eg due to overly restrictive
policies~\cite{OPENSSH_HANDCRAFTED_SECCOMP_HARD,
OPENSSH_HANDCRAFTED_SECCOMP_HARD_2}. Described in \S\ref{analysis-techniques},
these issues are a concern when deploying policies obtained manually or
dynamically, \emph{which is accepted by practitioners in
\Cref{tab:char-mainstream-soft}}.  Dunlap~\cite{Dunlap2022} covers this in
details.

\customparheader{Performance matters.}
Hardware historically imposed a heavy performance tax to compartmentalized
software. This explains why most of \Cref{tab:char-mainstream-soft} came
together with faster hardware in the 2000s.  A textbook case, the Windows NT
3.x kernel compartmentalized its graphics stack in the 90s under the influence
of microkernels, but soon reverted this in 4.0 due to
performance~\cite{NT_GDI_USERSPACE}. Still today, overheads are decisive when
shipping compartmentalization schemes to
production~\cite{RLBOX_PERFORMANCE_SHIP}. This is reflected in research, with a
majority of performance-focused works throughout \S\ref{sec:stateoftheart}.

\customparheader{Separation is effective but vulnerable.} Concrete impact on
bug exploitability has been documented where compartmentalization was pushed to
production -- most of \Cref{tab:char-mainstream-soft}. In the OpenBSD userland,
the OpenSSH and slaacd sandbox compartments successfully mitigated
code-execution flaws~\cite{OPENSSH_SANDBOX_MITIGATE,
OPENSSH_SANDBOX_UNEXPLOITABLE, SLAACD_SANDBOX_MITIGATE_1,
SLAACD_SANDBOX_MITIGATE_2}. Similar observations were made for Nginx
workers~\cite{NGINX_WORKER_MITIGATE_1, NGINX_WORKER_MITIGATE_2}, and web
browser site isolation.  Still, the protection is not limitless: interface
safety vulnerabilities were reported against
OpenSSH~\cite{OPENSSH_SANDBOX_ESCAPE, OPENSSH_SANDBOX_ESCAPE_2,
OPENSSH_SANDBOX_ESCAPE_3}, Firefox~\cite{Schumilo2022, FIREFOX_IPC_ISSUES},
Nginx~\cite{NGINX_WORKER_ESCAPE}, Chrome and generally site
isolation~\cite{CHROME_SITE_ISOLATION_BUG}.  These observations are more or
less direct effects of the ad-hoc nature of deployed compartmentalization
approaches. Although some works in \S\ref{sec:stateoftheart} are concerned with
the problem of interface safety, most scope it out to focus on performance.
This constitutes a gap between mainstream needs and research trends, which is
important: should compartmentalization become widespread, interface safety
flaws will be the main vulnerabilities of tomorrow.

\customparheader{PDMs are vastly manual.} For all of
\Cref{tab:char-mainstream-soft}, separation boundaries are manually identified
and maintained organically over time, a process costly in expertise and
efforts~\cite{OPENBSD_SEPPROCESS_1, OPENBSD_SEPPROCESS_2}.  As discussed in
\S\ref{automation-tradeoffs}, the engineering cost of manual separation limits
achievable separation granularities (for nearly all \totalfound apps,
separation is coarse with less than two to three domains), and makes separation less
approachable by the mainstream. Firefox library sandboxing, stemming from a
research project, is the only case of a non-fully manual
PDM~\cite{Narayan2020}. The fact that so few leverage automated methods may
also result from a mismatch between research goals and mainstream needs, as
research does not offer automated PDMs for data-centric separations (\cf
\S\ref{no-data-centric-sep}).

\customparheader{Diverse abstractions \& Focus on interfaces.} Abstractions
show a clear tendency towards sandboxing of untrusted code (\vs other models
from \S\ref{def:tms}). For the rest, abstractions feature an heterogeneous mix
of code- and data-centric designs, synchronous and asynchronous semantics, and
message-passing-based and shared-memory-based communication. Overall great attention
is dedicated towards interfaces. For instance OpenSSH leverages a custom
protocol with fully serialized and checked objects~\cite{Provos2003}, RLBox
leverages type data to systematically check and copy objects, and Nginx
leverages a very thin interface with nearly no communication from the untrusted
to the trusted world. The importance of interfaces is characterized by the
dominance of message-based approaches, which ease the checking of
interface-crossing data and control flows.

\customparheader{Importance of availability.} Most designs in
\Cref{tab:char-mainstream-soft} target a degree of availability. This comes in
contrast with research, which generally considers availability out of scope
(\S\ref{p2-taxonomy}). This may make it difficult to deploy many of the
previously described approaches under real-world expectations.

\customparheader{Mechanisms are page-table-centric.} \label{pt-centric-mechs}
Only two programs from \Cref{tab:char-mainstream-soft} do not rely on the page
table: Google V8, which leverages SFI, and Firefox library sandboxing, a
product of modern research~\cite{Narayan2020} which is mechanism-agnostic.  All
other designs strongly depend on the page table, an effect of their building
atop \customtt{fork()} semantics.  Regrettably, this dependency is hard to
break~\cite{Baumann2019}, making it difficult to reap the benefits of the
modern mechanisms we discuss in \S\ref{subsec:taxonomy-p3}: to leverage an
intra-address-space compartmentalization mechanism such as Intel MPK, program
code and data structures have to be redesigned to eliminate reliance on
\customtt{fork()}'s transparent address-space copy semantics.  Taking the
example of OpenSSH (\cf \Cref{tab:char-mainstream-soft}), which forks an
unprivileged process to perform its pre-authentication phase, this may require
substantial changes, e.g., identifying which data is required by the
unprivileged child, copying it explicitly, and refactoring the
pre-authentication phase code to function with the explicit copies.

%% file: 05-limitations.tex
\section{Outstanding Challenges}
\label{sec:limitations}

We conclude by consolidating the insights gained throughout this paper into
high-level challenges which we believe should be solved to mainstream modern
developments in compartmentalization and foster adoption.

\customparheaderb{Challenge 1: solving \ponetothree in isolation results in
unsuited solutions.}
Because software compartmentalization is such a large and complex problem,
\ponetothree are rarely, if ever, solved as a whole. Unfortunately, as we show
throughout this paper, this creates friction across the compartmentalization
stack as approaches are seldom composable: designing policies (or tools to
generate policies) which cannot be represented efficiently with existing
abstractions, abstractions which do not map to enforcement methods, or hardware
which does not efficiently enforce typical partitioning needs.

For example, most abstractions used in the mainstream are strongly tied to
processes and \customtt{fork()} semantics (\S\ref{sec:deployed}). This is a
problem as these semantics 1) do not compose, making it hard to use them in
conjunction with new compartmentalization abstractions, and 2) hinder the
ability to leverage new mechanisms, which are released at a fast pace
(\S\ref{subsec:taxonomy-p3}). Yet most \emph{new} abstractions proposed in
research still do not compose, and specialize towards specific mechanisms
(\S\ref{abstraction-specialization}). Will we repeat the mistake of
\customtt{fork()}~\cite{Baumann2019}, requiring each codebase to implement
several compartmentalization approaches? We need more concerted efforts across
the stack towards generic abstractions that compose and map to the ecosystem of
existing and future mechanisms.

Another instance of this problem is the performance cost of
compartmentalization.  It remains exceedingly hard to quantify the real
performance overhead of compartmentalization because research approaches it
narrowly: the focus on domain-switching latencies causes other less studied
costs such as those induced by mechanisms
(\S\ref{subsec:mechanism-performance}), the runtime costs of interface
protection (\S\ref{par:abstractionisafety}) and system call shielding
(\S\ref{p2-syscalls}), or of allocator hardening when heaps are shared, to be
left aside.  This calls for more efforts to better characterize the performance
costs of compartmentalization, and towards techniques and tooling to support
developers in estimating, diagnosing, and optimizing compartmentalization
performance costs early on and across \pone, \ptwo, and \pthree.

\customparheaderb{Challenge 2: creating and maintaining safe
compartmentalization policies is still too hard.}
The skills required to design safe policies are very specific: ensuring
interface safety, reasoning about the performance of compartmentalized software,
maintaining compartmentalizations over time; all constituting an art mastered by
trial and error.  Compartmentalization cannot go mainstream expecting
non-expert developers to acquire this art. As we show, this makes many
approaches described in \S\ref{p2-taxonomy} rather unsuitable for that purpose.
In fact, even when developers posses the required skills, compartmentalization
policies are still overly error-prone.  For instance, most of the works
described in \S\ref{p2-taxonomy} leave the job of securing internal and
external interfaces, a complex and particularly error-prone task, entirely to
the developer. As we observe in \S\ref{sec:deployed}, some software projects
have the skills, time, and budget to do so, but this is not the case of the vast
majority of the software ecosystem.

This calls for concerted efforts in two directions.  First, more work is needed
\emph{on approaches that do not require a policy from application developers}.
This can be approached with a focus on third-party dependencies that have the
skills and community, and where costs get amortized. Shared library and
software package APIs, for instance, should be designed from the ground up to
be transparently distrusted, following the example of, \eg V8
(\S\ref{sec:deployed}). This can also be approached through more works on
automated, generic compartmentalization, trading off security for deployability
(\S\ref{automation-tradeoffs}).  Second, more work is needed on
\emph{supporting the policy development process, for developers who have the
skills to do so}. There is a need for PDMs which can understand application and
boundary semantics; for more tooling to integrate compartmentalization into
long-term maintenance workflows to ensure safe separation over time; for more
automated interface safety checking methods; and for more fuzzing targeted at
the needs of compartmentalization.

\customparheaderb{Challenge 3: threat models are insufficiently challenged.}
Compartmentalization works largely make assumptions about interfaces
(adequately hardened, free of high-level logic bugs), compilers (no
correctness-security gap~\cite{DSilva2015, Xu2023}), shared components (libc,
threading libraries, memory allocators, are bug-free when shared), the kernel
(no confused deputies), or the hardware (the requirements of conditioned
mechanisms are satisfied, no side-channels).  \emph{These assumptions do not
hold in practice.}  Interfaces are porous and abstractions must be involved to
ensure safety by construction~\cite{Savage1994, Connor2020, Lefeuvre2023}
(\S\ref{subsec:abstractions-properties}).  Compilers~\cite{VanBulck2019,
Johnson2021, Johnson2023}, shared components and kernels~\cite{Connor2020,
Voulimeneas2022}, hardware~\cite{Oren2015, Lipp2018, Kocher2019, Canella2019}
(\S\ref{subsec:side-channels}) are all prone to separation-threatening flaws.
We need more offensive research exploring gaps in compartmentalization threat
models, characterizing and quantifying their impact, and defensive works with
holistic threat models to achieve truly systematic counter-measures.

\customparheaderb{Challenge 4: compartmentalization research deviates from the
needs of the mainstream.}
From analyzing the gap between research (\S\ref{sec:stateoftheart}) and
practice (\S\ref{sec:deployed}), we observe clear discrepancies between the focus of
compartmentalization in research and what practitioners run in production.
There are, for instance, very few works on availability in the
compartmentalization literature, yet this is what most deployed
compartmentalized programs target.  A similar observation can be made with
subject selections: there is no research work on PDMs for data-centric
compartmentalization, although it is popular among practitioners.  Research on
compartmentalization abstractions also developed a strong focus synchronous
semantics and shared memory in recent years, but practitioners take much more
diverse approaches mixing synchronous and asynchronous semantics, message
passing, and shared memory.  Lastly, research explores many hardware-specific
solutions, but practitioners need approaches that run on commodity hardware as
well, indicating that more efforts should be put into compartmentalization
schemes across \ponetothree that can deal with hardware heterogeneity.  This
broad gap between research and practical compartmentalization concludes this
SoK motivating for many research avenues in these under-explored areas.

%% file: 06-relatedworks.tex
\vspace{-0.05cm}
\section{Related Works}\label{sec:relatedworks}
\vspace{-0.05cm}

Shu at al.~\cite{Shu2016} surveys general isolation, and Acar at
al.~\cite{Acar2016} systematizes general Android security research, including
application compartmentalization.  Both have a wider scope than this SoK (\cf
\S\ref{def-comp}) and thus do not cover compartmentalization as systematically.
Both also predate many recent works: most of Tables
\ref{fig:taxonomy-algos}/\ref{fig:abstraction-taxonomy} appeared post-2016.

Other works cover vulnerability classes mitigated by compartmentalization such
as memory safety~\cite{Szekeres2013}, side-channels~\cite{Cauligi2022}, and
supply-chain~\cite{Ladisa2023} attacks, as well as mechanisms suitable for
compartmentalization~\cite{Cerdeira2020, Hu2021a, Jero2022}. These works are
orthogonal to this paper.  Others~\cite{Chen2007, Hu2015, Khandaker2020,
Lefeuvre2023, Lefeuvre2023a} classify interface safety issues and mitigations.
These complement this SoK, which draws a bigger picture of compartmentalization
challenges.  Sammler et al.~\cite{Sammler2019} models sandboxing to prove
safety properties.  These efforts motivate this SoK, confirming both the
validity as well as the limitations of their model.

%% file: 07-conclusion.tex
\vspace{-0.05cm}
\section{Conclusion}\label{sec:conclusion}
\vspace{-0.05cm}

Despite its benefits and decades of research and industry works,
compartmentalization remains a niche software engineering practice.  Through a
systematic study of \totalinstudy{} software compartmentalization works and
\totalfound{} deployed approaches, this paper sheds a light on the strengths
and limitations of current compartmentalization knowledge.  We stress that
popularizing software compartmentalization will require progress towards a more
holistic approach to compartmentalization; towards facilitating the definition
of compartmentalization policies; towards stronger and more holistic threat
models in the light of confused deputies and hardware limitations; as well as
towards more attention to the gaps between research and production approaches.

%% file: B-methodology.tex
\section*{A1. Taxonomy Appendix (complements \S\ref{sec:stateoftheart})}
\label{app:taxonomy-methodology}

\hypertarget{taxonomyappendix}{We include a paper if it concerns software
compartmentalization, as defined in \S\ref{def-comp}.}  We consider venues
ranked A* by CORE 2023 in security, systems, and programming languages: S\&P,
USENIX Security, CCS, NDSS, ASPLOS, OSDI, SOSP, PLDI.
\hypertarget{appsyssearch}{We manually triage titles and abstracts in an
\emph{ablative, conservative manner}}: when processing titles, papers clearly
unrelated to compartmentalization are discarded. If the title does not enable
an unambiguous decision, the paper is moved to the next stage (triage of
abstracts). Abstracts are sufficient to make an unambiguous decision in the
vast majority of cases.  We analyze the content of papers to determine if the
work address one of \ponetothree.

Researchers also published influential works pre-2003, in other (types of)
venues or communities, or in the industry.  We aim to cover these as well.
Since extending the systematic search to all these sources is untenable and
ultimately counter-productive, we address this with a recursive search in the
references called \emph{snowballing}~\cite{Wohlin14}, and by factoring in works
from our own expertise.  For the recursive reference search, we extract the
references of all \systematicsearch{} works identified previously
and repeat the filtering procedure.

%As a side note, although the general privilege separation literature before
%2003 is rich, we find that works that qualify as software compartmentalization
%are not numerous.

%% file: C-methodology.tex
\section*{A2. Source Search Appendix (complements \S\ref{sec:deployed})}
\label{app:source-search-methodology}

\vspace{-0.1cm}
\hypertarget{sourcesearchappendix}{The keywords considered} for the automated
search of package source-code in the Debian archive are: \emph{sandbox,
privilege, isolat*, separat*, compartment, partition, domain, capabilit*} (and
derived words, e.g., partition\emph{(ed$\vert{}$ing)},
separat\emph{(ed$\vert{}$ion)}). This yields 361 packages, which we
manually triage to determine if they implement compartmentalization,
resulting in \systemfound packages. The many false positives are composed of
software that do whole application sandboxing~\cite{Dunlap2022, SECCOMP}, drop
privileges (whole application least privilege), implement internal user
access-control policies that do not qualify as compartmentalization (\eg
databases), or use isolation or separation keywords to refer to other
development practices (\eg ``isolate a component in a class'', ``protected''
methods in C++). We look up the vendor website of each of the \systemfound
packages for software from the same vendor that also has a compartmentalized
design (missed in our initial search because less popular or not packaged in
Debian), resulting in \systemfoundvendor more compartmentalized programs, most
from the OpenBSD userland, totaling \totalsystemfound packages. We complete
this list with previous works (\S\ref{sec:stateoftheart}, \rworksfound programs),
and our knowledge of the field (\expertfound programs), to reach a corpus of
\totalfound mainstream compartmentalized programs.

\paragraph{Choice of the Debian Archive}

We choose the Debian archive because it 1) covers open-source software from all origins, 2) is very large (59K+
packages~\cite{DEBIAN_REPO}), and 3) comes with Popcon~\cite{POPCON}, a
popularity metric which maps to our ``mainstream'' criteria to
make the search practicable.

We recognize that a systematic search in the Debian archive does not cover
compartmentalized software in the mobile ecosystem (\eg Android or iOS). Beyond
the difficulty to integrate it in the paper's space, a systematic search of
mobile applications for compartmentalization patterns is difficult
due to the unavailability of sources in popular application stores.  We
therefore leave it for future works.

\paragraph{Program List}

\newcommand*{\systemns}{\textsuperscript{S}}
\newcommand*{\vendorns}{\textsuperscript{S}} % intentional same as \system
\newcommand*{\rworksns}{\textsuperscript{R}}
\newcommand*{\expertns}{\textsuperscript{K}}

\newcommand*{\system}{\systemns\xspace}
\newcommand*{\vendor}{\vendorns\xspace} % intentional same as \system
\newcommand*{\rworks}{\rworksns\xspace}
\newcommand*{\expert}{\expertns\xspace}

\emph{Labels.} From our systematic search: \systemns; From related works:
\rworksns; From our field knowledge: \expertns.

% TODO integrate Fuschia

% TODO add more microkernels here
% OpenBSD source 1: https://www.openbsd.org/innovations.html
% OpenBSD source 2: release notes, see scan_release_notes.sh
{
\small
\begin{itemize}[leftmargin=0.5cm,noitemsep]
\item \emph{OpenBSD Privilege-Separated Userland}: openssh\system, bgpd\vendor (openbgpd), dhclient\vendor, dhcpd\vendor, dvmrpd\vendor, eigrpd\vendor, file\vendor, httpd\vendor, iked\vendor, ldapd\vendor, ldpd\vendor, mountd\vendor, npppd\vendor, ntpd\system (openntpd), ospfd\vendor, ospf6d\vendor, pflogd\vendor, radiusd\vendor, relayd\vendor, ripd\vendor, script\vendor, smtpd\system (opensmtpd), syslogd\vendor, tcpdump\vendor, tmux\vendor, xconsole\vendor, xdm\vendor, Xserver\vendor (Xenocara), ypldap\vendor, pkg\_add\vendor, xlock\vendor, snmpd\vendor, dhcrelay\vendor, rbootd\vendor, pppoe\vendor, mopd\vendor, afsd\vendor, rdate\system (openrdate), sndiod\system, isakmpd\vendor, named\vendor, acme-client\vendor.
\item \emph{Web Servers}: Apache HTTPd\system, Nginx\system, Lighttpd\system
\item \emph{Browser Site Isolation}: Chrome\system, Firefox\system, Epiphany\system (all similar with implementation-related nuances)
\item \emph{Separated mail transfer agent architectures (and inspired approaches)}: qmail\expert, Postfix\system, djbdns\expert
\item \emph{Microkernels}: MINIX\rworks, L3/L4 family\rworks, and many others.
\item \emph{Single-application classes}: IRSSI\system, Debian man\system, DHCPCD\system, VSFTPD\system, Firefox library sandboxing\rworks, Google V8\system (libnode), Dovecot\system, Wireshark\system.
\end{itemize}
}